\documentclass[bjps]{imsart}
\RequirePackage{amsthm,amsmath,amsfonts,amssymb}
\RequirePackage[authoryear]{natbib}
\usepackage[dvipsnames]{xcolor}
\RequirePackage[colorlinks=true,citecolor=BlueViolet,urlcolor=BrickRed]{hyperref}
\usepackage{color}
\RequirePackage{graphicx}
\usepackage[utf8]{inputenc}
\usepackage[T1]{fontenc}
\usepackage{mathtools}
\usepackage{mathrsfs}
\usepackage{tabularray}
\usepackage{multicol}
\usepackage{booktabs}
\usepackage{enumerate}
\usepackage{enumitem}
\usepackage{cleveref}
\usepackage[many]{tcolorbox}

\definecolor{PaleBlue}{RGB}{0,163,243}

\tcbset{mystyle/.style={
 breakable,
 enhanced,
 outer arc=0pt,
 arc=0pt,
 colframe=PaleBlue,
 colback=PaleBlue!10,
 attach boxed title to top left,
 boxed title style={
 colback=PaleBlue,
 outer arc=0pt,
 arc=0pt,
 },
 title=TODO/COMMENT~\thetcbcounter,
 fonttitle=\sffamily
 }
}

\newtcolorbox[auto counter,number within=section]{todo}[1][]{
 mystyle
 }

\theoremstyle{plain}

\theoremstyle{remark}

\startlocaldefs

\DeclareMathOperator{\ddistGP}{\ensuremath{\mathsf{GP}}}
\DeclareMathOperator{\sign}{\mathrm{sign}} 
\DeclareMathOperator*{\argmin}{\mathrm{argmin}}
\DeclareMathOperator{\pr}{\Pr}
\newcommand{\dd}{\mathrm{d}} 
\newcommand{\calH}{\mathcal{H}}
\newcommand{\calP}{\mathcal{P}}
\newcommand{\calU}{\mathcal{U}}
\newcommand{\E}{\mathsf{E}}%

\endlocaldefs

\begin{document}

\begin{frontmatter}
\title{Choosing the Threshold in Extreme Value Analysis}
\runtitle{Choosing the threshold in extreme value analysis}
%
\begin{aug}
\author[a]{\inits{LRB}\fnms{L\'{e}o R.}~\snm{Belzile}\ead[label=e1]{leo.belzile@hec.ca}\orcid{0000-0002-9135-014X}}
\and
\author[b]{\inits{ACD}\fnms{Anthony C.}~\snm{Davison}\ead[label=e2]{anthony.davison@epfl.ch}\orcid{0000-0002-8537-6191}}
%
\address[a]{Department of Decision Sciences\\ HEC Montr\'eal\\ 3000, chemin de la C\^{o}te-Sainte-Catherine\\ Montr\'eal (Qu\'ebec)\\ Canada, H3T~2A7\\ \printead{e1}}
\address[b]{Institute of Mathematics\\ \'{E}cole polytechnique f\'{e}d\'{e}rale de Lausanne\\ EPFL-SB-PH, Station 8\\ 1015 Lausanne\\ Switzerland\\ \printead{e2}}

\runauthor{Belzile and Davison}

\end{aug}

\begin{abstract}
One of the two dominant approaches for univariate extreme value analysis is to model exceedances above a large threshold, the choice of which has a large impact on inference and whose uncertainty is often subsequently ignored. In this article we review more than 40 threshold selection procedures, including semiparametric methods based on Hill's estimator, visual diagnostics, goodness-of-fit tests, and others based on extended generalized Pareto models. Starting with the statistical properties underlying the various proposals, we provide a critical assessment of their strengths and weaknesses, discuss how they might be automated and describe the results of an extensive simulation study used to identify the most promising procedures. The approaches are compared using a long time series of daily rainfall totals from Padova.
\end{abstract}

\begin{keyword}
\kwd{generalized Pareto distribution}
\kwd{Padova rainfall series}
\kwd{peaks over threshold analysis}
\kwd{penultimate approximation}
\kwd{statistics of extremes}
\kwd{threshold selection}
\end{keyword}

\end{frontmatter}

\section{Introduction}

Natural catastrophes and economic disasters have impacted much of the globe over the past two decades, leading to a widespread appreciation of the potential impacts of rare events. A central role in the study of such events is played by extreme-value analysis, which employs statistical models to extrapolate from the data available to more severe, as-yet unseen, risks. The visibility of this domain of statistics has greatly increased, with a large community of researchers working on all of its aspects and mounting application of related methods.

Extreme-value theory is usually developed in terms of large events, typically analysed either as maxima of blocks of successive observations, or as exceedances above some high threshold. These approaches are linked by a unifying asymptotic framework that combines the theories of point processes for event times and of regular variation for event sizes. These limiting results provide asymptotic models that are fitted to non-asymptotic data, so an adequate fit of the models to the data is crucial to subsequent inferences, and in particular determines the trustworthiness of any extrapolation.

The initial theoretical results and applications of extreme-value theory considered block maxima \citep{fisher_tippett,Gnedenko:1943,Gumbel:1958}, but in the early 1970s attention turned to threshold exceedances \citep{Todorovic.Zelenhasic:1970,Todorovic.Rousselle:1971,Balkema.deHaan:1974,Pickands:1975}, and since the introduction of related statistical methods \citep{Davison:1984,Smith:1984,Davison.Smith:1990}, the so-called `peaks over threshold' (POT) approach has become widely applied. The key idea is that the extremes of data from a stationary time series be modelled by considering only those observations $Y$ that exceed a threshold $u$, since under mild conditions it can be shown that there exists a positive function $\sigma_u$ such that as $u$ approaches the upper support point of $Y$,
\begin{equation}\label{eq1}
\pr\{(Y-u)/\sigma_u>x\mid Y>u\} \to (1 + \xi x)_{+}^{-1/\xi}, \quad x>0,
\end{equation}
uniformly in $x$. Here $b_+=\max(b,0)$ and the parameter $\xi$ takes values in the real numbers; the right-hand side of \cref{eq1} is replaced by $\exp(-x)$ when $\xi=0$. This implies that an exceedance $X=Y-u$ over a sufficiently high threshold $u$ approximately follows the generalized Pareto distribution (GPD), i.e.,
\begin{equation}\label{eq2}
\Pr(X\leq x)\approx G(x) = \begin{cases} 1 - (1 + \xi x/\sigma_u)_+^{-1/\xi}, & \xi\neq 0,\\ 1-\exp(-x/\sigma_u),& \xi=0, \end{cases} \quad x> 0.
\end{equation}
Below we write $X\sim \ddistGP(\sigma_u, \xi)$ when $X$ has the distribution on the right-hand side of \cref{eq2}, and use $Y$ to denote general random variables. 

The distribution~\eqref{eq2} depends on two parameters. The shape parameter $\xi$ determines the upper tail weight of the density function, which is Pareto-like when $\xi>0$, exponential for $\xi=0$, and has support only in the interval $(0,-\sigma_u/\xi)$ when $\xi<0$. We shall see below that the scale parameter $\sigma_u$ depends on the chosen threshold $u$.

In the simplest applications a high but finite threshold $u$ is chosen using the available data, the parameters of \cref{eq2} are estimated from the $n_u$ exceedances over $u$, and this fitted model is used for extrapolation, often to values of $x$ that lie beyond the observed sample. The $(1-1/T)$-quantile of the distribution of annual maxima, or $T$-year return level $(T>1)$, is often the target of inference. In practice there may be numerous complications, most commonly that the series is non-stationary and that local dependence leads to the clustering of exceedances. In the original applied literature only the largest observation (the `peak') in each cluster was retained, but in line with current usage we shall use the abbreviation POT to refer to this entire approach to extremes.

There is a further trade-off between quality of fit to the available data and extrapolation beyond them. In \Cref{sec:basic}, we shall see that expression~\eqref{eq2} is rather flexible, and that ``penultimate'' approximations to the left-hand side of \cref{eq1} have the same form with different parameter values. This implies that a good fit to the available data should be balanced against the quality of extrapolation to higher, especially much higher, quantiles. It also suggests that different approaches to choosing the threshold $u$ may be appropriate when the purpose of an analysis is to estimate a particular parameter, such as a value-at-risk or an expectile, rather than to fit a model for subsequent general use.

The choice of threshold $u$ is critical to the successful application of POT methods, and the goal of this paper is to review methods of choosing it. Taking $u$ too high means that the number of exceedances $n_u$ is small, thus increasing the uncertainty of subsequent inferences, whereas taking $u$ too low may lead to poor extrapolation owing to the inclusion of non-extreme data. This trade-off has led to many suggestions for threshold selection, ranging from simple graphical procedures using stability properties of the GPD \citep[e.g.,][]{Davison.Smith:1990,Coles:2001} to sophisticated analytical procedures requiring stringent restrictions on the tail of $Y$. Certain procedures were compared numerically in \cite{Gomes.Oliveira:2001}, \citet{Schneider:2021}, \citet{Murphy.Tawn.Varty:2024} and elsewhere. The most comprehensive previous reviews of threshold selection are \cite{Scarrott.MacDonald:2012}, \citet{Caeiro.Gomes:2016} and \citet{Langousis:2016}. \cite{Scarrott.MacDonald:2012} cover the main ideas, with a focus on the splicing models described in~\Cref{sec:splicing}. Most of the methods covered by \cite{Caeiro.Gomes:2016} are semiparametric (cf. \Cref{sec:semihill}). Many further procedures for threshold selection have been proposed since these articles and book chapter appeared and are reviewed for the first time here.

We illustrate our discussion using daily rainfall data from Padova \citep{Marani.Zanetti:2015}. Exploratory analysis reveals strong seasonality, particularly for the number of rainy days, which form around 25\% of the total; we reduce the effect of such artefacts by taking only the months of July--September for the years 1878--2016, yielding $n=3311$ rainfall totals over 2mm. Although there are fewer rainy days during these summer months, the corresponding episodes of precipitation can be more extreme. For simplicity and since our objective is to illustrate and compare the results of threshold selection procedures, we treat the summer data as stationary.

The paper is organized as follows.
 In \Cref{sec:gev_gpd}, we sketch properties of the generalized Pareto distribution. and then in the following three sections describe threshold selection procedures. These can be broadly categorized based on the approach taken: threshold stability (\Cref{sec:visual}), extension of the GPD (\Cref{sec:extgp}), goodness-of-fit methods (\Cref{sec:gof}), or semiparametric procedures when $\xi>0$ (\Cref{sec:semihill}). \Cref{sec:numresults} outlines the results of extensive comparisons fully described in Appendices~\ref{sec:simulationstudy} and~\ref{sec:dataapp} and the results of applying the selection methods to the Padova series; other appendices give details for which there is insufficient space in the paper. \Cref{sec:discussion} contains a brief general discussion.
 

\section{Generalized Pareto distribution}\label{sec:gev_gpd}

%
%

\subsection{Basic notions}\label{sec:basic}

\subsubsection{First- and second-order conditions} \label{sec:secondorder}

The generalized Pareto limit of~\eqref{eq1} for rescaled threshold exceedances hold in wide generality. Let $Y$ be a random variable with distribution function $F$, define $F^{-}(p) = \inf\{y: F(y)\geq p\}$ for $p\in(0,1)$, and for $t>1$ let $b(t) = F^{-}(1-1/t)$. Then the so-called first order condition, which yields the generalized Pareto limit, holds if there exists a positive function $a$ such that \citep[Theorem~1.1.6]{deHaan.Ferreira:2006}
\begin{align*}
 \lim_{t \to \infty} \frac{b(ty)-b(t)}{a(t)} = \int_1^y s^{\xi-1}\dd s =\begin{cases}
(y^\xi-1)/\xi, & \xi \neq 0,\\\log y, & \xi=0,
 \end{cases} \qquad y > 0.
 \end{align*}
 A simple interpretation of the roles of $a(\cdot)$ and $\xi$ is possible if $F$ is twice differentiable and has cumulative hazard function $\calH(y) = -\log\{1-F(y)\}$ and reciprocal hazard function $r(y) = 1/\{\dd{\calH(y)}/\dd{y}\}$. Then we can take $a(t)=r\{b(t)\}$, write $\xi_t = r'\{b(t)\}$, and find that $\xi=\lim_{t\to\infty} \xi_t$.
 
 \citet{Smith:1987} shows that taking~\eqref{eq2} with $\xi_t$ rather than $\xi$ may give a better, so-called penultimate, approximation to the left-hand side of~\eqref{eq1}. A well-known case is the limiting exponential distribution for Gaussian exceedances, for which $\xi=0$, which is reached so slowly that for all practical purposes the shape parameter is negative \citep{fisher_tippett}, and this is what is found when fitting~\eqref{eq2} to Gaussian data. Such slow convergence might seem worrisome, but in practice the major issue is whether the GPD can provide useful extrapolations from data to levels likely to be used in applications, and of course this depends on the context. We illustrate this numerically in \Cref{sec:penultimate}.
Penultimate approximations can be related to further tail properties of $F$, which is 
said to be generalized regularly varying of second order at infinity if there exists a function $A(t)$ such that \citep[\S2.3]{deHaan.Ferreira:2006}
\begin{align}
H_{\xi, \rho}(y) &\coloneqq \lim_{t \to \infty} \frac{b(ty)-b(t)-a(t)\int_1^y s^{\xi-1} \dd s}{a(t)A(t)}
\\&= \int_1^y
s^{\xi-1}\int_1^s u^{\rho-1} \dd u \dd s,\ \qquad y>0, \rho \leq 0,\label{eqn:secondorder}
\end{align}
and such that $H_{\xi, \rho}(y) $ is not proportional to $\int_1^y s^{-\xi-1}\dd s$. The second-order auxiliary function $A(t)$ is regularly varying at infinity with tail index $\rho \leq 0$, i.e., $\lim_{t \to \infty} |{A(ty)}/{A(t)}| = y^\rho$ for $\rho \leq 0$, and also $\lim_{t \to \infty} A(t)=0$ with $A(t)$ ultimately not changing sign, so 
$\sign\{A(y_s)\}=\sign\{A(t+y_s)\}$ for some $y_s>0$ and for all $t>0$. If $b(t)$ is twice differentiable and $b'(t)$ is eventually positive, then \citep[Theorem~2.1]{deHaan.Resnick:1996} 
\begin{align}
 A(t) \coloneqq \frac{tb''(t)}{b'(t)}-\xi+1=r'\{b(t)\}-\xi = \xi_t - \xi; \label{2ndorderauxfun}
\end{align}
this is the difference between the penultimate and limiting values of the shape parameter.
\citet{Bucher.Zhou:2021} discuss these and similar conditions that apply to block maxima. 

Many methods mentioned in \Cref{sec:semihill} focus on the \cite{Hall.Welsh:1985} class of distributions, whose survival function admits the asymptotic expansion
\begin{align}
 1-F(y) = \alpha y^{-1/\xi}\left\{1 + \beta y^{\rho/\xi} + \mathrm{o}(y^{\rho/\xi})\right\}, \quad y\to\infty,
 \label{eq:HallWelshclass}
\end{align}
equivalent to a polynomial second-order auxiliary function $A(t) = \beta t^\rho$ for $\beta \neq 0$ and $\rho < 0$.

Although useful in theoretical work, the second-order functions such as $A$ or $H_{\xi,\rho}$ turn out to be extremely difficult to estimate, and it seems impossible to exploit this further structure in data analysis. 


\subsubsection{Threshold stability} \label{sec:tstabprop}
The GPD is threshold-stable: if $Y-u\sim \ddistGP(\sigma_u,\xi)$ and $v>u$ is such that $\pr(Y>v)>0$, then the conditional distribution of $Y-v$ given $Y>v$ is also GPD, with the same $\xi$ and scale parameter $\sigma_v=\sigma_u+ \xi(v-u)$. When $\xi<1$ we have $\E(Y-u) = \sigma_u/(1-\xi)$, so $\E(Y-v\mid Y>v) = \{\sigma_u + \xi(v-u)\}/(1-\xi)$. This is the basis for threshold stability plots, which display the empirical version of this conditional expectation and should be linear in $v$ if the GPD is adequate. The property also suggests diagnostics obtained by fitting the model over a grid of thresholds.

\subsubsection{Inhomogeneous Poisson process formulation}\label{sec:Pp}
The GPD can also be derived from a limiting Poisson process ${\mathcal P}$ under which events occur in the $(t,y)$-plane with measure
\begin{align*}
\Lambda[(t',t)\times[u,\infty)] = (t-t')\{1+\xi(u-\eta)/\tau\}^{-1/\xi}_+, \quad u\in\mathbb{R}, t>t', \quad \eta \in \mathbb{R}, \tau > 0. 
\end{align*}
The measure of the set ${\mathcal C}_u=[0,1]\times[u,\infty)$ under this model is 
\begin{align}\label{eq:mu}
 \mu(u) = \{1+\xi(u-\eta)/\tau\}^{-1/\xi}_+,
 \end{align}
 so the probability that the restriction of ${\mathcal P}$ to ${\mathcal C}_u$ has no points in ${\mathcal C}_{u+x}$ for $x>0$ is $\mu(x+u)/\mu(u) = (1+\xi x/\sigma_u)_+^{-1/\xi}$, where $\sigma_u = \tau + \xi(u-\eta)$, corresponding to \cref{eq1}. The vertical coordinates of the Poisson process ${\mathcal P}$ on ${\mathcal C}_{-\infty}$ can be generated by transforming a unit-rate Poisson process on the positive half line, $E_1, E_1+E_2, \ldots$, where the $E_j$ are independent standard exponential random variables, into the inhomogeneous process with points
\begin{align}\label{eq:Pp}
\eta + \frac{\tau}{\xi}\Bigg\{\Bigg(\sum_{j=1}^r E_j\Bigg)^{-\xi} - 1\Bigg\}, \quad r=1,2,\ldots,
\end{align}
which lie in the subset of the real line for which $\mu(x)>0$. Hence if we use an estimate of $\mu$ to transform the data to a unit-rate Poisson process, the choice of threshold amounts to choosing the smallest value of $u$ above which the transformed observations are consistent with this process.


\subsubsection{Martingale residuals}
The Markov properties of the order statistics $X_{(1)}\leq\cdots\leq X_{(n)}$ of a $\ddistGP(\sigma,\xi)$ random sample, which follow from the Poisson process construction described in~\Cref{sec:Pp}, allied to threshold stability, imply that the joint density of the order statistics can be written as
\begin{equation}
\label{eq:Renyi}
\prod_{j=2}^n f(x_{(j)} \mid x_{(j-1)}) f(x_{(1)})= \prod_{j=1}^n \frac{1}{\sigma_j}\left\{1 + \frac{\xi}{n+1-j}\frac{(x_{(j)}-x_{(j-1)})}{ \sigma_j}\right\}^{-(n+1-j)/\xi-1}_+,
\end{equation}
where $\sigma_j = (\sigma+\xi x_{j-1})/(n+1-j)$ and $x_{(0)}=0$ \citep[cf.][Appendix~A]{Oorschot.Segers.Zhou:2023}. This extends the \citet{Renyi:1953} representation for exponential data, as the smallest order statistic $X_{(1)} \sim\ddistGP(\sigma/n,\xi/n)$ and the increments $X_{(j)}-X_{(j-1)}$, conditional on $X_{(j-1)}=x_{(j-1)}$ have $ \ddistGP\{\sigma_j, \xi/(n+1-j)\}$ distributions, for $j=2, \ldots, n$, in addition to being conditionally independent of the lower order statistics. 
\cite{Stein:2023} explores using this representation to replace choosing the threshold by weighting the observations.

\subsection{Statistical properties of the generalized Pareto distribution}\label{sec:properties}

In this paper we consider the choice of threshold $u$, either within a fixed grid of ordered thresholds $u_1 < \cdots < u_J$ or within a subset of the order statistics $Y_{(1)}<\cdots <Y_{(n)}$ of a random sample $Y_1,\ldots, Y_n$; we write $\calU$ for the grid and call it fixed or random, respectively.

If $u$ is chosen, then we fit~\eqref{eq2} using the $n_u$ positive values of $Y_j-u$; $n_u$ is then random. Conversely we may choose $n_u$ and take $u=Y_{(n-n_u)}$ as the (random) threshold. 
As the limit in~\eqref{eq1} holds as $u$ increases to the upper support point $y_+$ of $Y$, we must use a so-called intermediate sequence for which $n_u/n\to 0$ as $n\to\infty$ in order for the estimator to be consistent, so that as $n_u$ increases the number of exceedances retained grows but comprises an increasingly small proportion of the original data. Simple rules such as using a fixed proportion of the data by setting $n_u\approx np$ for some fixed $p$ are inappropriate; we must rather use smaller fractions $n_u/n$ in larger samples, for example taking the $\lceil n- n^q\rceil$ largest order statistics, with $q=0.995$ or $q=0.999$. In practice a minimum number of observations is needed for fitting, whatever the size of the entire sample, so we must have $n_u\geq 20$, say.

\subsubsection{Maximum likelihood estimator} Maximum likelihood estimation makes no assumption on the sign of the shape parameter and is readily extended to more complex settings; the parameter estimators have large-sample normal distributions if \cref{eq1} holds and $\xi>-1/2$ \citep{smith}, though the normal approximation may be poor in small samples \citep{Suveges.Davison:2010}. Numerical problems can arise when $\xi<0$, because the support of the density, $\{x: \sigma_u + \xi x > 0\}$, then depends on the parameters, but these can be eased using
a reparametrisation due to R.~L.~Smith \citep[cf.][]{Davison:1984} that reduces the optimization to a unidimensional problem \citep{Grimshaw:1993} by
expressing the model in terms of $\xi$ and $\eta=-\xi/\sigma_u$. Apart from constants and when $\eta\neq 0$, its profile log likelihood,
\begin{align*}
 \ell_{\mathrm{p}}(\eta) = -n\log \left\{ -\dfrac{1}{n\eta}\sum_{i=1}^n\log(1-\eta x_i)\right\} -\sum_{i=1}^n \log(1-\eta x_i), \quad \eta < 1/x_{(n)},
\end{align*}
attains an overall maximum of $+\infty$ as $\eta\to 1/x_{(n)}$ and has a global maximum if the sample coefficient of variation exceeds unity and $\xi > 0$ \citep{delCastillo.Daoudi:2009}. 

The asymptotic covariance matrix of the maximum likelihood estimators $(\hat{\sigma}_u, \hat{\xi})$ depends on whether $u$ is fixed or an order statistic \citep[Remark~2.3]{Drees.Ferreira.deHaan:2004}, though $\hat\xi$ has asymptotic variance $(1+\xi)^2$ in both cases \citep[Corollary~2.1]{Drees.Ferreira.deHaan:2004}. These estimators can have large bias in small samples.

\subsubsection{Hill's estimator} \label{sec:hillest}

The well-known \cite{Hill:1975} estimator of a positive shape parameter $\xi$ averages the log spacings of the $n_u$ largest order statistics of a random sample relative to the random threshold $u=Y_{(n-n_u)}$, assumed to be positive:
\begin{align}
 H_{n,n_u}
 = \frac{1}{n_u}\sum_{i=1}^{n_u}\left\{ \log Y_{(n-i+1)} - \log Y_{(n-n_u)} \right\}. \label{eq:Hillest}
\end{align}

If the upper tail is Pareto-like ($\xi>0$), the \cite{Weissman:1978} estimator of the quantile at level $1-p$ is $Q_{n_u}(1-p) = Y_{(n-n_u)}\{n_u/(pn)\}^{H_{n,n_u}}$. Although $ H_{n,n_u} $ is consistent for $\xi>0$ for an intermediate sequence $n_u$, it is not location-invariant. Less widely-used estimators for real-valued $\xi$ include the moment estimator of \cite{Dekkers.deHaan:1989}, those based on generalized median, trimmed mean and mean excess functionals of \cite{Beirlant.Vynckier.Teugels:1996b}, and a location-scale invariant estimator based on an extreme $U$-statistic \citep{Oorschot.Segers.Zhou:2023}; the last uses log spacings and thus is undefined in the presence of ties. For the latter, as well as for the shape estimator of \cite{Wager:2014}, more work is needed to derive accompanying estimators of scale and quantiles, along with confidence intervals.

%
 A sufficient condition for asymptotic normality of Hill's estimator is second order regular variation (cf.~\Cref{sec:secondorder}) with $\lim_{n_u \to \infty}{n_u}^{1/2}A(n/n_u) = \lambda\in \mathbb{R}$, subject to which \citep[Theorem 3.2.5]{deHaan.Ferreira:2006}
 \begin{align}
n_u^{1/2}(H_{n,n_u} - \xi) \to \mathsf{normal}\{\lambda/(1-\rho), \xi^2\}, \quad \rho \leq 0. \label{eq:asympnormalityHill}
\end{align}
The asymptotic bias of $\lambda/(1-\rho)$, which is unknown in applications, is thus dictated by the rate at which the
number of extreme observations grows relative to the total sample size and by the function $A(t)$, which is distribution-specific. Hill's estimator has a smaller asymptotic variance than the maximum likelihood estimator, but is restricted to positive values of $\xi$.

\subsubsection{\texorpdfstring{$L$-moments}{L-moments}} \label{sec:lmom}
Probability weighted moments are often used for estimation in hydrological applications. For a generalized Pareto variable $X$ with $\xi<1$ we can define $\alpha_r=\mathsf{E}[X\{1-F(X)\}^r]= \sigma/\{(1+r)(1+r - \xi)\}$ ($r=0,1,\ldots$), yielding estimators based on the identities $\xi = -\alpha_0/(\alpha_0-2\alpha_1) + 2$ and $\sigma=2\alpha_0\alpha_1/(\alpha_0-2\alpha_1)$ with the $\alpha_r$ replaced by empirical counterparts \citep[][eq.~5]{Hosking.Wallis:1987}. Such estimators are linear in the observations, and thus have good small-sample properties. They are closely relate to $L$-moments \citep{Hosking:1990}, which are linear combinations of probability weighted moments of the form
\begin{align*}
\lambda_s = \sum_{r=0}^{s-1} (-1)^{s-1-r}\binom{s-1}{r}\binom{s-1+r}{r}\alpha_r, \quad s=1,2,\ldots. 
\end{align*}
Another estimator based on the $L$-skew, $\tau_3 = \lambda_3/\lambda_2$, takes $\xi_{\text{lmom}} =(3 \tau_3 - 1) / (1 + \tau_3)$ and $\sigma_{\text{lmom}}(1 - \xi_{\text{lmom}}) (2 - \xi_{\text{lmom}}) \lambda_2$, where $(\sigma_{\text{lmom}}, \xi_{\text{lmom}})$ are obtained by replacing probability weighted moments by their unbiased estimators.

\subsubsection{\texorpdfstring{Padova data}{Padova data}} \label{sec:PDdata}
\begin{figure}[t!]
\centering
\includegraphics[width=\textwidth]{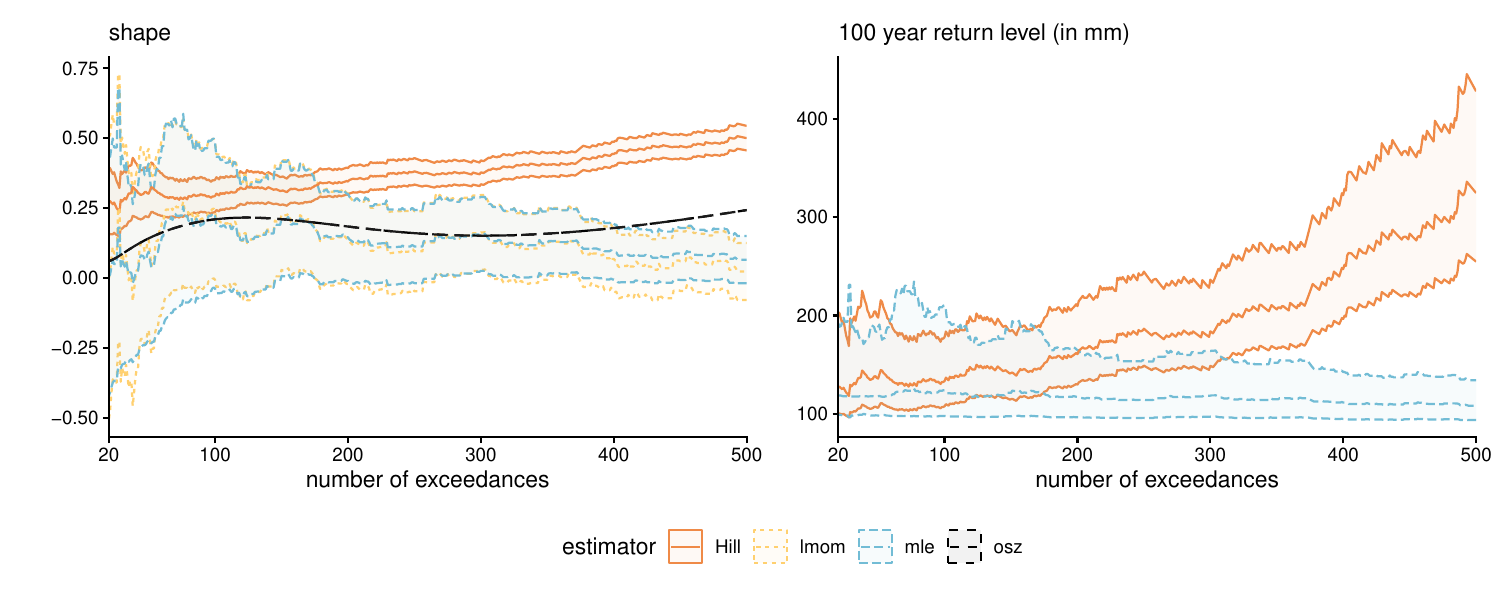}
\caption{Variation with the threshold of shape parameter estimates (left) and 100-year return levels (right) for different estimators: the maximum likelihood and (\textsc{mle}) $L$-moments (lmom) \citep{Hosking.Wallis:1987} of the generalized Pareto parameters, the \cite{Hill:1975} estimator, and the $U$-statistic estimator of \cite{Oorschot.Segers.Zhou:2023} (osz). Shape estimates are shown with Wald-based 95\% confidence intervals. The \textsc{mle} and $L$-moments return levels are based on the GPD quantile function and with 95\% profile-likelihood confidence intervals, ignoring variation due to the random number of exceedances, whereas the Hill estimator uses the \cite{Weissman:1978} quantile estimator with the saddlepoint confidence interval estimator of \cite{Buitendag.Beirlant.deWet:2020}.}
\label{figthstab}
\end{figure}

\Cref{figthstab} illustrates the variability of shape estimates obtained using different estimators applied to the Padova data. The $U$-statistic-based estimator of \cite{Oorschot.Segers.Zhou:2023} varies smoothly, but the others are sensitive to the threshold. Estimators from nearby thresholds are strongly correlated. The Hill estimator yields very large return levels as $n_u$ increases, due to the extrapolation with large shape values. The maximum likelihood and $L$-moment estimator (not shown) are rather stable and near identical, both close to the results of a Bayesian analysis.

In most applications good estimation of $\sigma_u$ and $\xi$ is not the goal of analysis, and a key aspect is stability of extrapolation to risk functionals such as high quantiles. \Cref{figthstab} shows that, while the maximum likelihood estimates of the shape parameter are rather variable, the quantile estimates are more or less constant, whereas those based on the Weissman estimator increase steadily as the number $n_u$ of exceedances grows.

\section{Methods based on stability} \label{sec:visual}
\subsection{Moment stability plots}
Threshold stability (\Cref{sec:tstabprop}) underpins various graphical diagnostics. If $X_1, \ldots, X_n$ follow $\ddistGP(\sigma, \xi)$ above a threshold $u$, and $\xi<1$, then
\begin{align}
e(v) = \mathrm{E}(X - v \mid X>v) = \frac{\sigma_u - \xi u}{1-\xi} + \frac{\xi}{1-\xi} v, \quad v\geq u. \label{eq:mrl}
\end{align}
This motivates the mean excess, or mean residual life, plot \citep[cf.][\S~2]{Davison.Smith:1990}, which graphs the empirical mean excesses $\hat e(v) = n_v^{-1}\sum_{i=1}^{n_v} (X_i-v)_+$ against thresholds $v$ in a grid $ \calU$, in the hope that the graph will become roughly straight above some threshold. Its interpretation is aided by supplementing the graph with uncertainty bands.

\citet[][Section~2.2]{Langousis:2016} proposed automating this informal procedure by taking order statistics as candidate thresholds. Their idea is that for each $u \in \calU= \{ Y_{(1)} \leq \cdots \leq Y_{(n-20)}\}$, a weighted linear regression should be fitted to pairs $\{Y_{(i)}, \hat e(Y_{(i)})\}$ $(i=1, \ldots, n-10)$, with $j$th largest order statistic weighted by the reciprocal sample variance of exceedances over $u$. The procedure returns the threshold that minimises the weighted mean squared error. Discarding the ten largest order statistics ensures that the weights are not too large, as can arise if there are (near) ties in the largest observations.

The left-hand panel of \Cref{fig:padova_tstab} shows the mean residual life plot for the Padova data with $\calU$ consisting of quantiles at levels 0.8 (13.58mm) to 0.995 (65.49mm) in increments of 0.005; there are 663 and 17 exceedances at the limiting values of $u$. The plot seems to stabilize towards 50mm, but the \cite{Langousis:2016} procedure returns the lowest candidate threshold, despite the fitted regression line lying below all observations above 40mm.

\subsection{Parameter stability plots}
Another popular visual procedure is the parameter stability plot \citep{Davison.Smith:1990}. If the generalized Pareto model holds above $u$, then the shape $\xi$ and/or the modified scale $\sigma_{v}-\xi v$ are constant for $v\geq u$, suggesting that we check for stability of point estimates for values of $v$ in a grid $\calU$, with these estimates supplemented by appropriate confidence intervals. Such plots can be difficult to interpret because the estimates are based on overlapping samples and do not account for penultimate effects. Moreover they provide no assurance that the model fits adequately, so they should be complemented with diagnostics of fit focused on the upper tail. In an attempt to automate such diagnostics by mimicking common practice, in our simulation study we return the lowest threshold $u$ for which the shape parameter estimate at all higher thresholds falls within the confidence interval of $\xi$ at $u$. Unless there are strong penultimate effects, this leads to the selection of very low thresholds.

 The parameter stability plot in the central panel of \Cref{fig:padova_tstab} gives much lower shape estimates than Hill's estimator. There is an initial increase of $\hat{\xi}$ until $u=40$ or so, but the substantial uncertainty implies that the smallest candidate threshold would be adequate.
%

\subsection{Hill plots}
Hill plots graph the Hill estimator $H_{n,n_u}$ of \cref{eq:Hillest} against the number of order statistics $n_u$ used for estimation. The usual recommendation is to choose a number of order statistics $n_u$ in a region where the Hill estimator stabilizes based on visual inspection of pairs $(n_u, H_{n,n_u})$ for different values of $n_u$. This is notoriously difficult because the sample paths of the process $\{n_u, H_{n,n_u}\}$ are analogous to those of a modified Brownian motion \citep{Mason.Turova:1994} and can be very rough; see \Cref{figthstab}. The estimates are best graphed against the log number of exceedances and may be smoothed using a moving window estimator. \citet[\S\S4.4--4.6]{Resnick:2006} describes alternative Hill plots, including ones in which $H_{n,n_u}$ is replaced by a local average. \cite{Danielsson:2019} suggest automating the selection of $n_u$ by looking for a drop in the variance of the Hill estimator. They use the proportion of the $m$ subsequent tail indexes, say $1/H_{n,n_u+1}, \ldots, 1/H_{n,n_u+m}$, that lie at most $\varepsilon$ from $1/H_{n,n_u}$, suggesting setting $\varepsilon = 0.3$ and taking the largest value of $n_u$ such that 90\% of the observations are within the bound. This does not account for the correlation between estimates, and the fact that the variance of the tail index depends on both $\xi$ and $n_u$.

\begin{figure}[t!]
 \centering
\includegraphics[width = \textwidth]{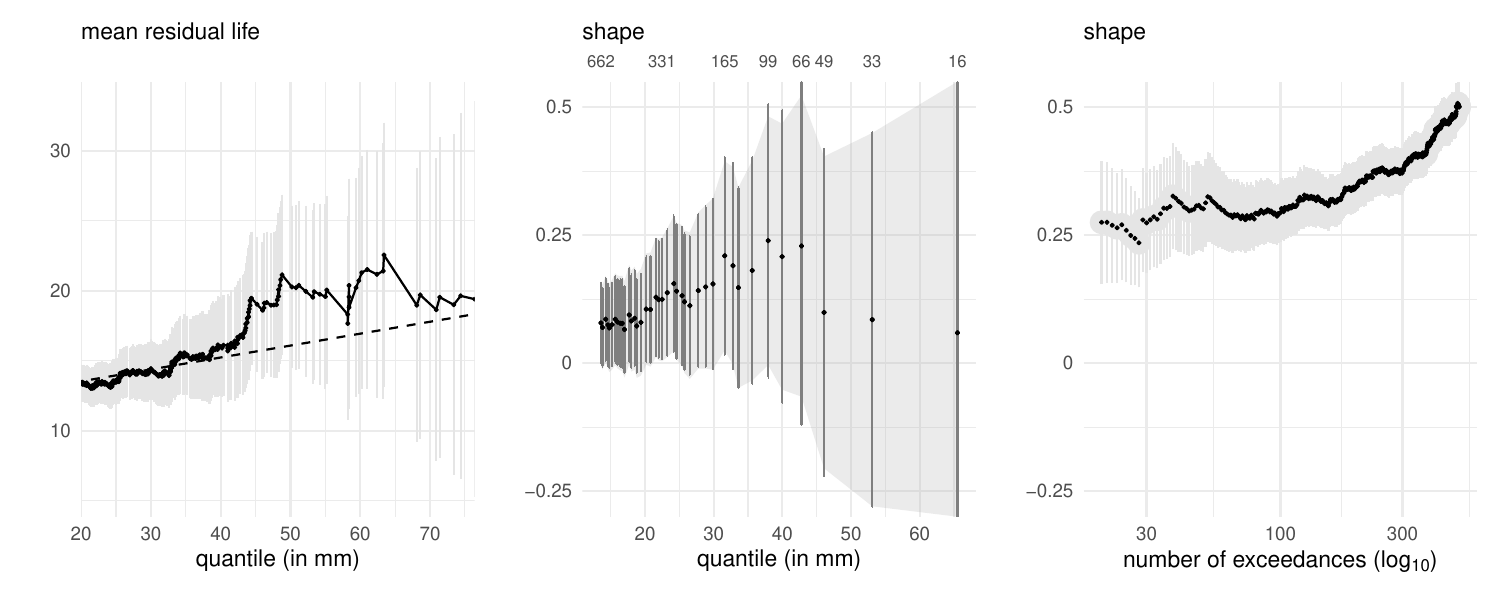}
\caption{Stability plots for the Padova summer rainfall data, with point estimates as a function of the threshold or the number of exeedances, and 95\% Wald pontwise (line segment) and simultaneous (shaded area) confidence intervals. Left: mean residual life from \cref{eq:mrl}, with weighted regression line. Center: stability plot for the shape parameter with pointwise (line) and simultaneous intervals; number of exceedances for selected threshold is shown on top. Right: Hill estimates of the shape parameter against the $\log_{10}$ of the number of exceedances.}
\label{fig:padova_tstab}
 \end{figure}

The right-hand panel of \Cref{fig:padova_tstab} shows the Hill plot, graphed as recommended against $\log_{10} n_u$, for $20 \le n_u < 663$. Its estimates of $\xi$ are much higher, and seem stable for $n_u\lesssim 150$. The result of the \cite{Danielsson:2019} procedure depends on how the range of order statistics is restricted, returning a threshold at 47.6mm with $\hat{\xi} = 0.29$, and $n_u=54$ if all the non-zero rainfall data are retained, but giving $n_u=20$ and $\hat{\xi}=0.28$ above the $0.995$ quantile, if $n_u$ is restricted to be less than 1000.

\subsection{Likelihood-based procedures} \label{sec:likstab}

Stability diagnostics can be based on the large-sample distribution of the maximum likelihood estimator. \cite{Thompson:2009} suggest performing Pearson's test of normality for differences of the successive estimators $\hat{\sigma}_j - \hat{\xi}_j u_j$ of the scale parameter of the Poisson process model for thresholds $u_j$ in a fixed grid $\calU$, stopping when the hypothesis is rejected at level $\alpha = 0.2$. If the model applies, then these differences should have mean zero, but their correlation complicates the assessment of significance for the tests.

\cite{Wadsworth:2016} supposes that Poisson processes of the form described in \Cref{sec:Pp} apply with possibly different parameters on each interval of $\calU$, and shows that, if the asymptotic regime with a common shape parameter has been reached, differences of the estimators $\hat \xi_1,\ldots, \hat\xi_J$ based on the exceedances of consecutive thresholds can be rescaled to form asymptotically independent standard normal variables, $\varepsilon_j=(\hat{\xi}_{j+1}-\hat{\xi}_{j})/\{(I^{-1}_{j+1}-I^{-1}_j)_{\xi,\xi}^{1/2}\}$, for $j=1,\ldots, J-1$; here $I_{j}$ is the Fisher information matrix for exceedances of threshold $u_j$, and the subscript $(\xi,\xi)$ denotes the corresponding matrix element. Her idea is that this standardisation to Gaussian white noise should be effective within the asymptotic regime, but that the $\varepsilon_j$ can be treated as Gaussian white noise with another mean and variance at thresholds below this regime. Thus a likelihood ratio test for a changepoint in the distribution of $\varepsilon_1,\ldots, \varepsilon_{J-1}$ should indicate the start of the asymptotic regime, as shown by the lowest threshold above which the $p$-value for the test exceeds some level $\alpha$. Experience suggests that this procedure is fragile; in particular the difference in inverse Fisher information matrices is often not positive definite, and altering $\calU$ can lead to completely different conclusions. The left-hand panel of \Cref{fig2} shows the \citet{Wadsworth:2016} white-noise sequence $\varepsilon_j$, which suggests picking the lowest possible threshold for the Padova data.

 \section{Extended models} \label{sec:extgp}

The fitting of asymptotic models such as the generalized Pareto distribution to a small subset of the available data, sometimes called the \emph{extremal paradigm},\ is the dominant approach to the modelling of rare events, but it has drawbacks. One is that the sharp division of the data into `extreme' and `bulk' at a threshold is unrealistic. A smoother transition towards the tail might be achieved by modifying the generalized Pareto model so that its upper-tail properties remain satisfied in the limit but it is more flexible near the threshold, which could then be lowered without serious misspecification. The goal is to balance potential gains in precision of estimation due to including more data against increased uncertainty due to estimating additional parameters.
In this section we describe the main approaches to extending the generalized Pareto distribution and discuss corresponding threshold choice procedures. 

\subsection{Piecewise generalized Pareto models} \label{method:changepoint}

In Section~\ref{sec:secondorder} we saw that the penultimate shape parameter typically depends on the threshold. \cite{Wadsworth.Tawn:2012} propose approximating this behaviour using a two-phase inhomogeneous Poisson process with different measures of the form of~\cref{eq:mu} on intervals $(u_1, u_2]$ and $(u_2, +\infty)$. The intensity functions are constrained to be continuous at $u_2$, and tests for equality of the two shape parameters are performed at each pair $(u_1,u_2)$ in a fixed grid. The models can be hard to fit, and the procedure entails multiple testing.

\begin{figure}[t!]
\centering
\includegraphics[width=\textwidth]{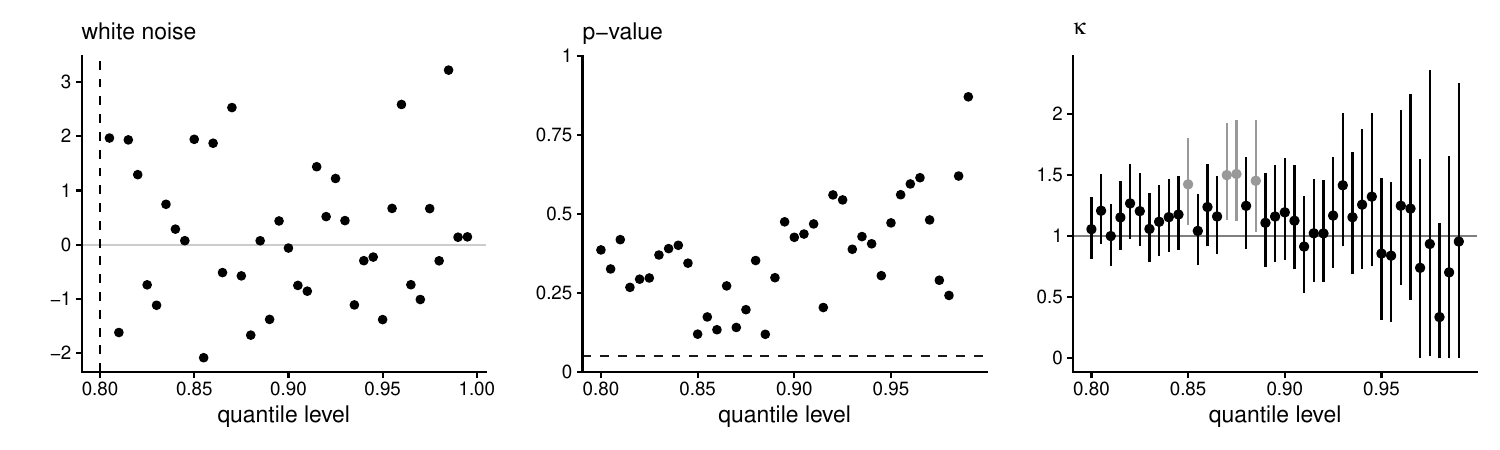}
\caption[Threshold selection diagnostics for Padova rainfall.]{Threshold selection diagnostics for the Padova rainfall series. Left: white noise sequence diagnostic of \cite{Wadsworth:2016} from \Cref{sec:likstab}; rescaled differences of shape parameter estimates, $\{\xi^*_i\}$. The dashed vertical line indicates the lowest threshold at which we fail to reject the hypothesis that the sequence is standard Gaussian white noise. Center: $p$-values of the \cite{Northrop.Coleman:2014} score test against threshold. The horizontal dashed line at $\alpha = 5\%$ gives the cutoff for individual tests. Right: parameter stability plot for $\kappa$ for the beta extended generalized Pareto model \citep{Gamet.Jalbert:2022}, with pointwise profile-based 95\% confidence intervals. Intervals that fail to cover $\kappa=1$ are in grey.}
\label{fig2}
\end{figure}

\cite{Northrop.Coleman:2014} instead propose fitting a piecewise generalized Pareto distribution, $\mathsf{PGP}(\sigma, \xi_1, \ldots, \xi_J)$, which consists of interval-truncated GPDs with potentially different shape parameters on the intervals defined by a fixed grid $\calU\cup\{+\infty\}$. Continuity of the density function at $u_2,\ldots, u_J$ imposes $J-1$ restrictions on the scale parameters, so the full model has only $J+1$ parameters. The model can be challenging to fit, so the authors propose score tests of the hypotheses that $\xi_j=\cdots=\xi_J$, which correspond to a single generalized Pareto model above $u_j$, against the alternative that $\xi_j,\ldots, \xi_J$ are different. They plot the corresponding $p$-values against the thresholds, and for  a given significance level $\alpha$ then choose the lowest threshold with a non-significant $p$-value or at which the $p$-values for all higher thresholds are all non-significant. The central panel of \Cref{fig2}, which shows this plot for the Padova data, leads to the choice of the $0.8$ quantile as threshold. 

Under the original Northrop--Coleman procedure, the rejection probability at any particular threshold is $\alpha$, but the tests are dependent because the data are re-used. Moreover, its power depends on $\calU$ and especially on $u_J$. Neither of their proposed approaches to choosing a threshold controls the overall error rate: the sequence of $p$-values does not allow comparison of the null models across thresholds because the data differ for each test. An alternative procedure tests a $\mathsf{PGP}(\sigma, \xi_1, \ldots, \xi_J)$ model with $\xi_j = \cdots = \xi_J$ for $j=1,\ldots, J-1$ against the full piecewise model using likelihood ratio statistics with $J-j$ degrees of freedom. As the data do not change, this also allows model comparison using information criteria. This entails full fits of $J$ models, first fitting the ordinary generalized Pareto model to all the data above $u_1$, then successively fitting the models with $\xi_1\neq \cdots\neq\xi_{j}= \cdots= \xi_J$ for $j=2,\ldots, J$, using the final parameter values for one fit as starting-values for the next; this is more stable and faster than direct fitting of the full model. The likelihood will increase when each parameter is added, but the full model remains quite difficult to fit and the shape parameter estimates can vary appreciably between successive intervals.  

\subsection{Extended generalized Pareto models}

Starting from \citet{Papastathopoulos.Tawn:2013}, numerous attempts have been made to embed the GP distribution function $G(x;\sigma,\xi)$ in a more flexible model without changing its limiting tail behaviour. Apart from the hope of being able to use a lower threshold for inference on the upper tail, or even entirely avoiding choosing a threshold, this approach may also allow the simultaneous modelling of extremes in both tails \citep{Naveau:2016}. A general construction of such models rests on a continuous distribution function $F_\kappa$ on $[0,1]$, whose density is $f_\kappa$. The EGP$(\sigma,\xi, G_\kappa)$ distribution function is then \citep{Naveau:2025}
\begin{align*}
\Pr(X\leq x) = F_\kappa\{ G(x;\sigma,\xi)\},
\end{align*}
and the corresponding density and quantile functions are 
\begin{align*}
f(x;\sigma,\xi) f_\kappa\{ G(x;\sigma,\xi)\}, \qquad
\sigma Q_\xi\left\{f_\kappa^{-1}(p)\right\}, \qquad 0<p<1, \end{align*}
where $Q_\xi(p) = \{(1-p)^{-\xi}-1\}/\xi$ for $\xi\neq 0$ and $Q_0(p) = -\log(1-p)$ is the generalized Pareto quantile function.

Extended generalized Pareto distributions provide more flexibility for modelling departures from the limiting form, and thus may allow a user to choose lower thresholds and increase the sample size, while allowing the additional parameters to capture departures from a generalized Pareto tail for lowest exceedances. In the case of scalar $\kappa$, with $\kappa=a$ retrieving the generalized Pareto model, parameter stability plots can be used to find a region in which $\kappa \approx a$ and the shape parameter stabilizes. Hypothesis tests for $\kappa=a$ can also be constructed, although testing will only be regular if $a$ lies inside the parameter space. However, estimators of $\kappa$ are typically strongly correlated with those of $(\sigma, \xi)$, so parameter uncertainty balloons, even if that for return levels increases less sharply.
\cite{Huet:2026} proposed fitting an extended generalized Pareto model to (a subset of the) data and picking the largest point above which the density is convex as threshold (the largest zero of the second derivative of the density $\mathrm{arg\ \!max}_{x}\partial^2 f(x; \boldsymbol{\theta})/\partial x^2=0.$ This is justified by the fact that the generalized Pareto density is strictly convex when $\xi>-1/2$, and both models should behave similarly in the tail. As EGPD models are appropriate only for positive data, this procedure anyway requires the set of exceedances of an initial threshold.

The \cite{Papastathopoulos.Tawn:2013} models have prescribed behaviour at $x=0$: for example, if $F_\kappa=u^\kappa$ for $\kappa>0$, the density at the origin is $g(0;\sigma,\xi) = 1/\sigma$ but $f_\kappa(0)$ equals zero if $\kappa>1$ or $+\infty$ if $\kappa<1$. \cite{Gamet.Jalbert:2022} propose two further models that avoid this behaviour, but only their extended beta model leads to regular asymptotics for comparisons with the generalized Pareto. \cite{Stein:2021} considers other constructions.

 Parameter stability plots for $\kappa$ for the Padova data and the \cite{Gamet.Jalbert:2022} beta model, shown in the right-hand panel of \Cref{fig2}, suggest no evidence against the generalized Pareto model from the 0.88 quantile onwards, perhaps due to a lack of power. The estimator of $\kappa$ is highly correlated with that of $\xi$, which leads to wide confidence intervals and complicates assessment. The 100-year return levels from fitting different extended generalized Pareto models to these data at different thresholds yield essentially the same inference, although the upper bound of the confidence interval associated with the EGP model tends to be higher, most likely owing to the added uncertainty due to expanding the model class.  The two extended generalized Pareto models, the beta model of \cite{Gamet.Jalbert:2022} and the power model of \cite{Papastathopoulos.Tawn:2013} exhibit similar behaviour for the sampling distribution of the return level estimator. The \cite{Huet:2026} procedure returns a threshold of zero based on the entire Padova rainfall series, as the EGP fit is convex over the whole domain, with $\hat{\xi}\approx 0.25$; depending on the lower bound considered for fitting the EGP, we can however obtain very different thresholds.

%
%

%

\subsection{Mixture and splicing models} \label{sec:splicing}
The generalized Pareto model is only valid for exceedances above the threshold $u$, so it is tempting to splice it together with a specification for the bulk of the data, which lies below $u$. Section~6 of \cite{Scarrott.MacDonald:2012} and the discussion in \cite{Hu.Scarrott:2018} outline the ideas in the many papers that take this approach; the latter also provides advice and a software implementation. These methods typically do not consider threshold selection, so are not compared in our simulations, but for completeness we sketch them here.

The distribution function of a ``bulk model-based tail fraction'' splicing model has disjoint components below and above $u$ with GPD $G(x; \sigma_u, \xi)$ \citep{Hu.Scarrott:2018},
\begin{align*}
F(y; \boldsymbol{\theta}) =
\begin{cases}
H(y; \boldsymbol{\theta}), & y \leq u, \\
H(u; \boldsymbol{\theta}) + \left\{1-H(u; \boldsymbol{\theta})\right\} G(y-u; \sigma_u, \xi)
,& y > u,
\end{cases}
\end{align*}
where $H$ is the distribution function of the bulk, whereas a ``parametrized tail fraction'' model uses a mixture of truncated distributions with mixing probability $\phi_u$,
\begin{align*}
F(y; \boldsymbol{\theta}) =
\begin{cases}
(1-\phi_u){H(y; \boldsymbol{\theta})}/{H(u; \boldsymbol{\theta})} ,& y \leq u, \\
(1-\phi_u) + \phi_u G(y-u, \sigma_u, \xi),& y > u.
\end{cases}
\end{align*}
The second specification is more flexible and its likelihood factorises, so it is preferable unless the bulk model fits well: this can be assessed using standard diagnostic tools.

\begin{figure}[t!]
\centering
\includegraphics[width=\textwidth]{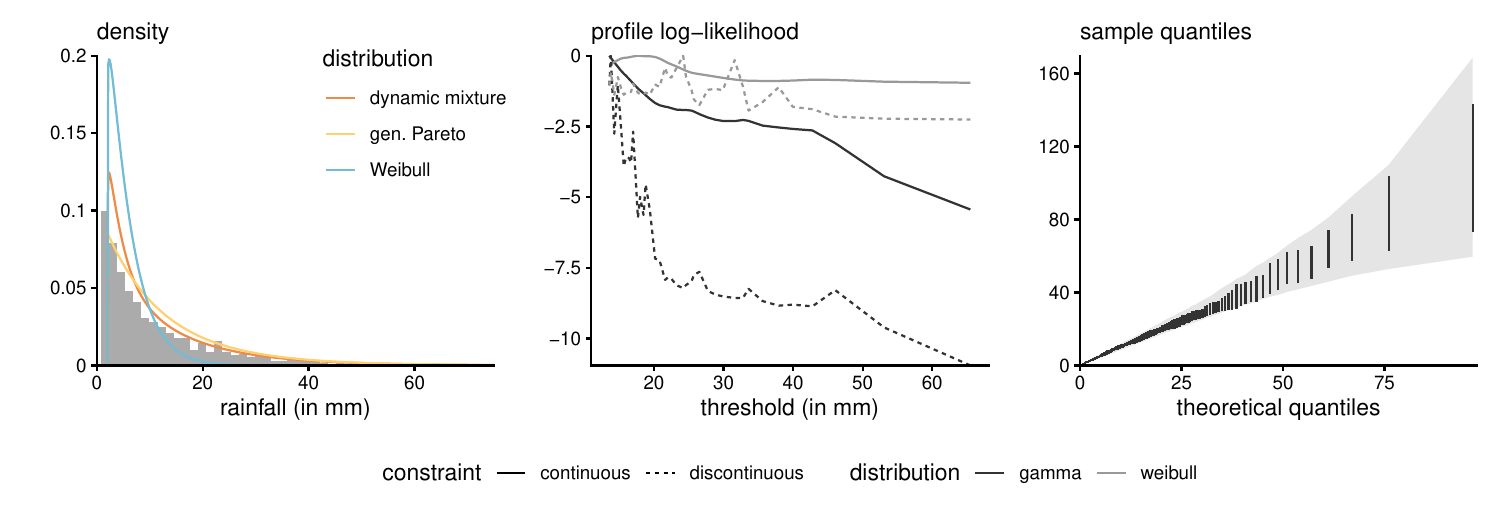}
\caption[Splicing and dynamic mixture models for Padova data]{Splicing and dynamic mixture models for the Padova data, and threshold selection diagnostics. Left: dynamic Weibull-generalized Pareto mixture density of \cite{Frigessi.Haug.Rue:2002}. Center: profile log-likelihood functions for the threshold parameter $u$ for Weibull and gamma splicing models, with and without continuity constraints at the threshold. Right: quantile-quantile plot for the \cite{Murphy.Tawn.Varty:2024} procedure at the selected threshold of $u=13.58$mm, corresponding to the 0.8 quantile, with tolerance bands (grey) and segments giving the approximate positions of observations, all of which fall within the bands. }
\label{fig4}
\end{figure}

Splicing models describe the whole distribution, so users can treat $u$ as a parameter and choose it to give the best fit; its uncertainty can also be taken into account. On the other hand the resulting estimate may be unduly influenced by the bulk model, so robustness of the fit to the choice of $u$ is critical. Goodness-of-fit measures, such as quantile-quantile plots for both parts of the distribution, are often helpful. Profile likelihoods for $u$ can sometimes be plotted, but they are typically uninformative and multimodal: the central panel of \Cref{fig4} shows profile log-likelihood functions for the threshold based on various mixture models for the Padova data, which, though unstable, all point to a low threshold. Using the bulk model-based rather than the parametrized tail fraction does not change the overall picture, but the profiles are smooth when continuity is imposed at $u$. Neither Weibull (threshold at 0.855 quantile) nor gamma (threshold at quantile 0.8) models fit the bulk adequately, but quantile-quantile plots (not shown) show satisfactory fit in the tails, with shape estimates of respectively $0.08$ and $0.16$ for the GP component.

Continuity of parametric models can be imposed by equating the density, and perhaps its derivatives, at the threshold. This leads to more natural-looking models, but the resulting interaction between the bulk and tail models can lead to poor quantile estimation if either part fits badly. \cite{Carreau.Bengio:2009} consider a Gaussian-generalized Pareto splicing model, imposing continuity on the density and its derivative at $u$, which is treated as a free parameter defined implicitly through the constraints or by constrained optimisation. \cite{RiveraMancia:2014} consider similar smoothness constraints and a Bayesian nonparametric mixture for the generalized Pareto tail model, while \cite{MacDonald:2011} use kernel density estimation with a continuity constraint for the bulk, with the kernel bandwidth chosen using a cross-validated likelihood. \cite{doNascimento.Gamerman.Lopes:2012} consider a discontinuous model in the Bayesian paradigm and argue that the discontinuity has no impact for posterior predictive inference about the tail if $u$ is treated as random.

Many papers have considered Bayesian estimation, as careful prior specification can help deal with multimodal likelihood functions and capture the uncertainty in threshold selection. For example, \cite{Behrens.Lopes.Gamerman:2004} considered splicing models where $u$ is treated as a model parameter, with a left-truncated distribution centered around high quantiles as its prior, and \cite{Bermudez:2001} confine the threshold to a reasonable range by using a doubly-truncated Poisson prior for the number of exceedances. \cite{Tancredi.Anderson.OHagan:2006} used the inhomogeneous point process parametrization to reduce dependence on the threshold.

 Splicing mixtures in which the bulk is nonparametrically specified, using a Dirichlet process mixture of gamma densities, a mixture of uniform densities or kernel density model, are more flexible than their parametric counterparts, but are more computationally demanding and convergence can be problematic; Chapter~5 of \cite{MacDonald:thesis} considers the use of Tukey's sensitivity curves for maximum likelihood estimators of the extremal mixture parameter, when transferring data between bulk and tail.

Splicing and extended generalized Pareto distributions can be combined by smoothly blending the bulk and tail through a transition function that allows the weight in the mixture density to vary dynamically \citep{Frigessi.Haug.Rue:2002}, with 
 \begin{align*}
 f(y) \propto \{1-w(y; \boldsymbol{\vartheta})\}h(y; \boldsymbol{\theta}) + w(y; \boldsymbol{\vartheta}) g(y; \sigma, \xi),
 \end{align*}
and the weighting function $w(\cdot)$ taken to be a location-scale Cauchy distribution function. Here the bulk and tail models are defined over the whole real line, with the tail given by a dynamic weighted mixture. Since the parameters of $w$ govern the mixing weight, there is no guarantee that the generalized Pareto model will dominate in the upper tail, and this approach tends to perform poorly in practice. \cite{Scarrott.MacDonald:2012} remark that a transition function with a high density in a small neighborhood yields sharp transitions. The left-hand panel of \Cref{fig4} shows the fit of the \citet{Frigessi.Haug.Rue:2002} dynamic mixture model to the Padova rainfall totals greater than 2mm. The generalized Pareto shape parameter takes the reasonable value of $0.06$, but the tail model contributes 50\% of the weight from the minimum exceedance onwards, indicating possible contamination of the tail by the bulk.

\section{Goodness-of-fit measures} \label{sec:gof}

Threshold stability plots do not measure the fit of the tail model to the data, making it hard to assess whether inference and subsequent extrapolation are reliable. Another natural approach is to base threshold selection on the fit of a generalized Pareto model to the tail data.

\subsection{Quantile-quantile plot distance}

The fit of data to a distribution $F_0$ is often assessed using quantile-quantile {(Q-Q)} plots of ordered data against plotting positions, typically $\{p_j = j/(n+1): j = 1, \ldots, n\}$, that have been transformed using the quantile function $Q_0$ of $F_0$. When the shape of $F_0$ depends on parameters, as with the GPD, uncertainty can be assessed by parametric simulation from a fitted distribution $\hat F_0$ \citep[\S4.2.4]{Davison.Hinkley:1997}. The model is refitted to each simulated sample and the corresponding quantile function is evaluated at the plotting positions.

%
%

To avoid visual inspection of Q-Q plots for a grid of thresholds, \cite{Varty.Tawn:2021} suggest using a metric such as (weighted) mean absolute or mean square error for agreement of the empirical and theoretical positions. \cite{Murphy.Tawn.Varty:2024} suggest fixing grids $\calU$ of thresholds and $\calP = \{p_1, \ldots, p_m\}$ of plotting points at which to evaluate the fit and using a bootstrap scheme for $u\in\calU$:
\begin{itemize}
 \item for $b=1,\ldots, B$, generate bootstrap samples of exceedances of $u$ and obtain the corresponding fitted GP models $\hat F_0^{(b)}$;
 \item obtain the $x$-axis positions using the generalized Pareto quantile function for $\hat F_0^{(b)}$, evaluated at $\calP$;
 \item obtain the $y$-axis positions from the empirical quantile function evaluated at $\mathcal{P}$;
 \item compute the average metric over $\mathcal{P}$ and the $B$ bootstrap samples.
\end{itemize}
The threshold $u$ giving the smallest average metric is chosen. By modifying the sampling scheme this proposal can be used with censored or non-identically distributed data, and for time-varying thresholds, but it requires an optimization for each bootstrap sample and each threshold. \cite{Varty.Tawn:2021} propose using a nonparametric bootstrap and plotting on the exponential scale with $l_1$ norm as metric, whereas \cite{Murphy.Tawn.Varty:2024} compute the metric on the generalized Pareto scale. The latter gives lower mean squared error for quantile estimators when the shape $\xi$ is positive, since the exponential model downweights larger observations and leads to the choice of lower thresholds. The TAILS approach of \cite{Collings:2025}, who suggest assigning more weight to large values of $\mathcal{P}$ and discarding those below a user-selected cutoff $p_0$, leads to the selection of much higher thresholds. 

\cite{Murphy.Tawn.Varty:2024} report simulation studies that suggest reduced root mean squared error for their quantile estimators relative to the procedures of \cite{Northrop.Attalides.Jonathan:2017} and \citet{Wadsworth:2016}. Using this with the $l_1$ norm on the Padova data returns the lowest candidate threshold; the quantile-quantile plot in the right-hand panel of \Cref{fig4} shows a good fit, but very wide uncertainty for the highest quantiles.

\subsection{Minimum distance selection}

A popular approach for fitting to network data exhibiting power-law upper tails \citep{Clauset.Shalizi.Newman:2009} treats observations above $u$ as an exact sample from a Pareto distribution with survival function $(x/u)^{-1/\xi}$ on $[u, \infty)$.
They suggest picking the threshold $u$ to minimize the Kolmogorov--Smirnov distance between the empirical distribution function of the exceedances, $\hat{F}_{n_u}$ and the best fitting Pareto model obtained using Hill's estimator, and assess uncertainty using a nonparametric bootstrap.
\cite{Drees:2020} 
consider candidate thresholds among the order statistics and show that the resulting shape parameter estimator has a higher asymptotic mean squared error than Hill's estimator and that the chosen thresholds tend to be higher than would be optimum. A related proposal of \cite{Danielsson:2019} suggests picking $n_u$ to minimize the supremum distance between \cite{Weissman:1978}'s quantile estimator and the empirical quantile of the largest 
exceedances,
but simulations of \cite{Murphy.Tawn.Varty:2024} suggest that this has high bias that leads to overly high thresholds, due both to the chosen criterion and to not accounting for the variability of the order statistics. For Pareto tails with $\xi>0$, \cite{Goegebeur.Beirlant.deWet:2008} also mention choosing $n_u$ to minimize a weighted Cram\'er--von Mises statistic between log exceedances and the fitted exponential quantiles.

\subsection{Bootstrap procedure}
To account for variation in the number of exceedances above a fixed threshold $u$, one can employ a  bootstrap procedure. We estimate as usual the parameters $\sigma_u, \xi$ of the generalized Pareto distribution from \cref{eq2} from the $n_u$ exceedances of a threshold $u$, and consider the mixture distribution function consisting of point mass at observed values and a generalized Pareto tail
\begin{align}
\widetilde{F}(y) = \frac{1}{n}\sum_{i=1}^{n-n_u}\mathbf{1}_{Y_{(i)} \leq y}+ \mathbf{1}_{u < y}\frac{n_u}{n
} G(y-u, \hat{\sigma}_u, \hat{\xi}).
\label{eq:semipar}
\end{align}
A bootstrap sample is drawn by taking $n$ uniform draws and inverting \cref{eq:semipar}, leading to varying numbers of exceedances above $u$, to which the generalized Pareto model is fitted.

\cite{Caers.Beirlant.Maes:1999} suggest minimizing the mean squared error of the shape parameter, which they estimate from the discrepancies between the resulting bootstrap shape estimates and the original estimate value $\hat{\xi}$. In practice, the variance tends to dominate the bias, so this approach leads to low thresholds. \cite{Gonzalo.Olmo:2004} propose checking the fit of the generalized Pareto model using a Kolmogorov--Smirnov statistic that compares the empirical distribution function and fitted GPD, generalizing \cite{Pickands:1975}, and to use the bootstrap based on~\cref{eq:semipar} to approximate its sampling distribution.

\subsection{\texorpdfstring{Methods based on $L$-moments}{Methods based on L-moments}}

The scale-invariance of $L$-moment ratios, $\tau_r =\lambda_r/\lambda_2$ ($r=3, \ldots$) (\Cref{sec:lmom}), can be used to evaluate distributional fit. \cite{SilvaLomba.FragaAlves:2020} propose automating visual selection from plots of these ratios by noting that the $L$-skewness $\tau_3=(1+\xi)/(3-\xi)$ of the generalized Pareto distribution is related to its $L$-kurtosis, $\tau_4 = f(\tau_3) = \tau_3(1+5\tau_3)/(5+\tau_3)$. Their idea is to compare their empirical estimates ($\hat{\tau}_3, \hat{\tau}_4)$ and the implied $L$-moments $\{\hat{\tau}_3, f(\hat{\tau}_3)\}$ on a plot of $\tau_4$ against $\tau_3$ for $u\in\calU$, then choose the threshold $u$ to minimises the $l_2$ distance to the theoretical curve, i.e., taking
\begin{align*}
\hat u = \adjustlimits\argmin_u\min_{\tau_3} \{\hat{\tau}_3(u) - \tau_3\}^2 + \{\hat{\tau}_4(u) - f(\tau_3)\}^2.
\end{align*}

\cite{Kiran.Srinivas:2021} likewise propose fitting the generalized Pareto distribution to the exceedances of $u$, mapping them using the probability integral transform to the unit exponential scale and calculating the first $L$-moment and $L$-skewness, $(\hat{\lambda}_1, \hat{\tau}_3)$, from the transformed sample. They then use Monte Carlo sampling from unit exponential or analytical approximation to estimate the mean and variance matrix of these estimators, and return the threshold that minimises the Mahalanobis distance between $(\hat{\lambda}_1, \hat{\tau}_3)$ and the ``theoretical'' values for the unit exponential model, found by simulation.

\cite{Solari:2017} use the weighted Anderson--Darling statistic of \cite{Sinclair.Spurr.Ahmad:1990},
\begin{align*}
A_R^2(\boldsymbol{y})=\frac{n_u}{2}-\sum_{i=1}^{n_u} \left\{\frac{2n_u-2i+1}{n_u}\log(1-z_i)+2z_i \right\},
 \end{align*}
for each candidate threshold $u$ and corresponding set of  $n_u$ exceedances, where $z_i = G(y_{(n_u+1-i)}-u; \hat{\sigma}_u, \hat{\xi})$,
 with $G$ the distribution function of \cref{eq2} using the $L$-moment estimates $(\hat{\sigma}_u, \hat{\xi})$. The null distribution of $A_R^2$ is approximated by parametric bootstrap simulation. \cite{Solari:2017} recommend choosing the threshold for which the $p$-value for $A_R^2$ is highest, but this would lead to random selection if the data were exactly generalized Pareto.

\subsection{Bayesian predictive measures} \label{sec:bayesiansurprise}
\cite{Lee:2015} propose a threshold stability plot that shows the Bayesian $p$-value for a summary statistic that captures the agreement or `surprise' between sample and data simulated from the posterior distribution. Too low a threshold will be suggested by departures from the expected value of 0.5 for $p$-values, which are uniform if the GPD fits well. Samples from the posterior distribution of the generalized Pareto parameters $(\sigma, \xi)$ are then used to compute the test statistic for the observed exceedances and simulated posterior predictive samples. \cite{Lee:2015} make box-and-whisker plots of $p$-values, so the procedure must be replicated to provide decent Monte Carlo estimates of the variability. The results are noisy even with data simulated from the null model, making it hard to diagnose lack of fit. Furthermore, there is some arbitrariness in the choice of goodness-of-fit statistic. The partial posterior predictive distribution is based on the posterior conditional on the statistic, and is equivalent to a leave-one-out scheme if the statistic considered is the posterior log density of the $r$th order statistic. This choice yields overly-variable $p$-values, visible in simulations of \cite{Lee:2015}. More work  seems to be needed on appropriate measures of surprise and to reduce the computational burden.

\begin{figure}[t!]
\centering
\includegraphics[width=\textwidth]{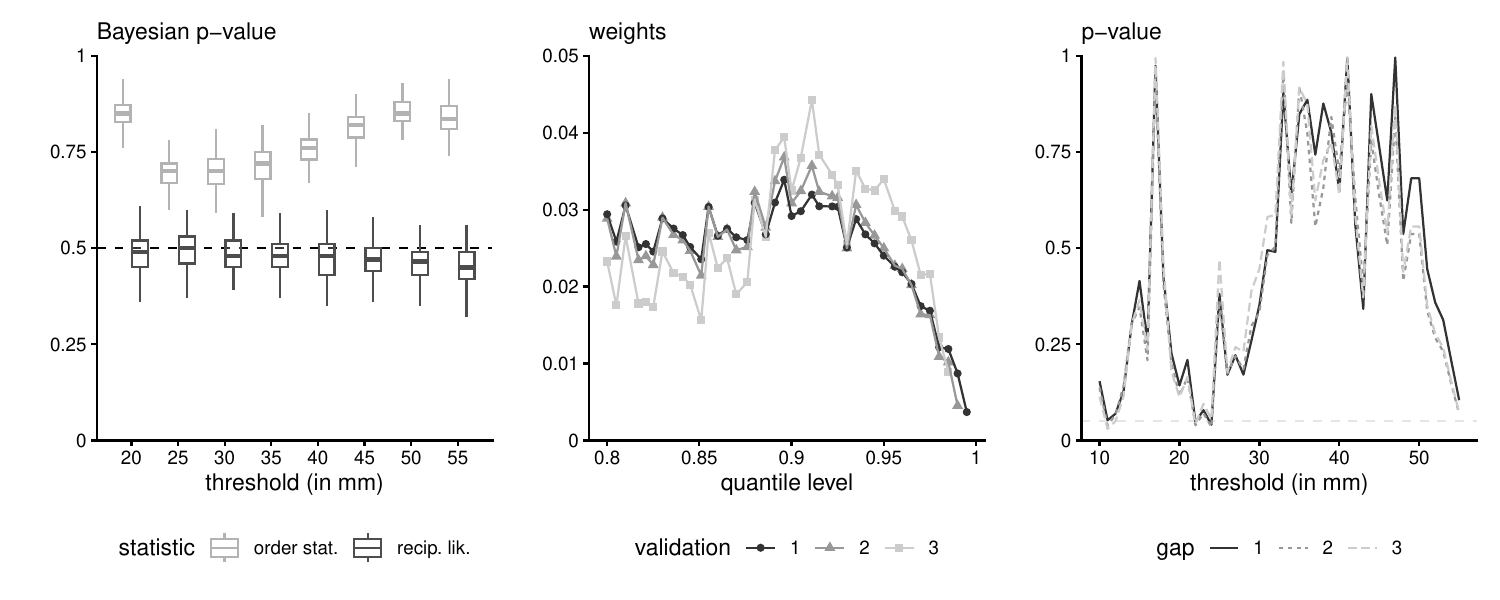}
 \caption[Threshold selection diagnostics for Padova rainfall (3)]{Threshold selection diagnostics for the Padova rainfall series. Left: \cite{Lee:2015} measures of surprise (\Cref{sec:bayesiansurprise}) with two test statistics, the reciprocal likelihood $1/f(\boldsymbol{x}; \boldsymbol{\theta})$ or equivalently $-\ell(\boldsymbol{\theta}; \boldsymbol{x})$ and the distribution of the largest order statistic, $X_{(n)}$. The box-and-whiskers plots are based on 100 replications at each threshold. Center: model averaging weights obtained by Bayesian leave-one cross-validation using the top order statistics as validation threshold \citep{Northrop.Attalides.Jonathan:2017}. Right: $p$-values for the information matrix test of \cite{Suveges.Davison:2010} as a function of the maximum gap length $K=1,\ldots, 3$.}
\label{fig3}
\end{figure}

The left-hand panel of \Cref{fig3} shows measures of surprise for the Padova data using the likelihood of the largest order statistic as validation test statistic and simulating samples from the partial posterior predictive distribution, and using the the reciprocal likelihood based on the full posterior. The former suggests that all thresholds are unsuitable, the latter that the lowest possible threshold would be suitable. It is unclear whether the reciprocal likelihood lacks power to detect poor fit.

\cite{Northrop.Attalides.Jonathan:2017} propose a Bayesian method based on leave-one-out cross-validation with a binomial-generalized Pareto (BGP) model, whose density for $\tau_u = \Pr(Y > u)$ is
\begin{align*}
f_u(y) = (1-\tau_u)^{\mathbf{1}_{y \leq u}}\{\tau_u g(y-u; \sigma_u, \xi)\}^{\mathbf{1}_{y > u}}.
\end{align*}
 They use a fixed grid $\calU$ and a validation threshold $v > u_J$ above which they assess model performance: the BGP model above $v$ follows from threshold-stability arguments, with corresponding likelihood contribution $f_v$. The `cross-validation predictive density' for the left-out observation $x_r$ from $\boldsymbol{x}_{-r}=\{x_1, \ldots, x_{n_{u_j}}\} \setminus \{x_r\}$, the full sample of exceedances minus the $r$th observation exceeding candidate threshold $u_j$, is
\begin{align*}
 f_v(x_r \mid \boldsymbol{x}_{-r}, u_j) = \int f_v(x_r \mid \boldsymbol{\theta}, \boldsymbol{x}_{-r}) p_j(\boldsymbol{\theta} \mid \boldsymbol{x}_{-r}) \mathrm{d} \boldsymbol{\theta},
\end{align*}
with $p_j$ the posterior density of the parameters at threshold $u_j$. The cross-validation predictive density, $f_v(x_r \mid \boldsymbol{x}_{-r}, u_j)$, is approximated using Monte Carlo samples from $p_i(\boldsymbol{\theta} \mid \boldsymbol{x}_{-r})$ obtained via importance sampling from the full posterior ${p_j(\boldsymbol{\theta} \mid \boldsymbol{x})}$. The proposed measure of fit, an estimate of the negative Kullback--Leibler divergence,
\begin{align*}
 \hat{T}_v(u_j) = \sum_{r=1}^{n_{u_j}} \log \hat{f}_v(x_r \mid \boldsymbol{x}_{-r}, u_j),
\end{align*}
uses all exceedances to assess fit, unlike in \cite{Lee:2015}. The selected threshold maximizes this diagnostic over $\calU$. \cite{Northrop.Attalides.Jonathan:2017} use Bayesian model averaging to account for the uncertainty due to threshold selection, with the prior predictive distribution replaced by the proxy $\exp\{\hat{T}_v(u_j)\}$; weights as a function of the threshold are shown in the central panel of \Cref{fig3} for validation based on up to the three largest candidate thresholds. The approach is computationally-intensive but the choice of threshold is automatic and is based on the fit of the model. The BGP likelihood allows the comparison of different thresholds, but does not provide a valid joint distribution for the whole sample above the lowest threshold; this seems to impact comparisons of predictive fit. Simulations in \cite{Murphy.Tawn.Varty:2024} indicate that the method leads to highly variable threshold choices, and we concur with this.

\subsection{Distributional tests} \label{sec:sequnivtest}

One way to automate threshold selection is to fit a generalized Pareto distribution at each threshold in a fixed grid $\calU$, compute a goodness-of-fit statistic, and then perform tests sequentially for $u\in\calU$ until rejection. %
%
%
%
\cite{Choulakian.Stephens:2001} considered Anderson--Darling and Cramér--von Mises statistics for testing whether the data above a threshold are generalized Pareto. They tabulated the null distributions of these statistics, which do not depend on the scale parameter, though \cite{Langousis:2016} point out that these distributions are valid only for unrounded data and that their validity suffers further under serial correlation. \cite{Bader.Yan.Zhang:2018} proposed an improved approach using Monte Carlo simulation to approximate the null distribution for more values of the shape parameter, with linear extrapolation for the largest values of the test statistic. Although invalid in this setting, ForwardStop \citep{GSell:2016} is used in a bid to control the false positive rate and thus to alleviate multiple testing concerns. \cite{Bader.Yan.Zhang:2018} suggest using the $p$-values for a goodness-of-fit statistic at each threshold in a fixed grid $\calU$ to form the sequence $D_j= -j^{-1} \sum_{i=1}^j \log(1-P_i)$ $(j=1, \ldots, J)$ and choosing the largest threshold for which $D_j < \alpha$. Simulation suggests that the Anderson--Darling statistic gives higher power than other statistics considered, and despite the lack of theoretical guarantees for the control of the false discovery rate, the size distortion of the procedure seems to be small. Simulations in their paper suggest that the approach leads to lower mean squared error for prediction of extreme quantiles than the competing methods considered. Adjustments are needed for the null distribution approximation to be reliable with rounded data.

\subsection{Time series}
Most results so far assume independent data, but clustering of extremes and serial dependence are widespread in applications. Under mild conditions, \cite{Ferro/Segers:2003} show that the intervals between successive exceedances of a high threshold $u$ of data from a stationary sequence $Y_1, \ldots, Y_T$ approximately follow a mixture of an exponential distribution with rate $\theta\lambda_u$ and a point mass at zero; here $\lambda_u$ is the rate of exceedances over $u$ and the extremal index $\theta\in(0,1]$ is the reciprocal of the limiting mean cluster size. Under conditions on extremal clustering, \cite{Suveges.Davison:2010} adapt this result and for data with $n_u$ exceedances and intervals $T_1, \ldots , T_{n_u-1}$ define the set of truncated intervals $S(u,K)=\{\max\{T_{i}(u)-K, 0\}: i=1,\ldots, n_u-1\}$ for possible gaps $K=0, 1, \ldots$, and the corresponding log-likelihood function for $\theta$, $\ell\{\theta; S(u,K)\}$. If the model is well-specified, then for every threshold $u_j$ in a grid $\calU$ and gap $K$, the Fisher information $I(\theta) =- \mathrm{E}[\ell''\{\theta; S(u_j,K)\}]$ equals the variance of the score statistic $J(\theta) = \mathrm{Va}[\ell'\{\theta, S(u_j, K)\}]$. Results of \cite{White:1982} suggest testing whether $D(\theta) = J(\theta) - I(\theta)$ equals zero, with $J$ and $I$ replaced by their empirical counterparts evaluated at the maximum likelihood estimator and the test statistic compared to its null asymptotic distribution. This procedure gives a $p$-value for each combination $(u_j, K)$, and leads to multiple testing. \cite{Fukutome:2015} suggest an \textit{ad hoc} automation that keeps all pairs for which the test statistic is less than 0.05 (i.e., the $p$-value exceeds 0.82) and choosing the combination of run length and threshold that yields the most clusters; they recommend having at least 80 exceedances at every threshold, but this does not palliate the multiple testing. The application of this to the Padova data in the right-hand panel of \Cref{fig3} shows relative insensitivity to $K$, and suggests that the model misspecification is not significant beyond 25 mm, giving $n_u=271$.

\section{Semiparametric methods} \label{sec:semihill}

Threshold selection for moment estimators is typically based on minimizing an asymptotic mean squared error for Hill's estimator,~\cref{eq:Hillest}.
Related methods rely on assumptions about the second-order behaviour of the survival function and the sign of the shape parameter, and may perform poorly in finite samples. Most do not account for the noise due to the insertion of point estimators into formulae, and numerical implementations may not respect the need for a minimum number of upper order statistics, $n-n_u$, for estimation to be reliable, so users must check the results with care.

\subsection{Expansions for the asymptotic mean squared error}

\cite{Caeiro.Gomes:2016} review, but do not numerically compare, semiparametric procedures for positive shape parameters for distributions that are second-order regularly varying; see our Simulation study~2. These procedures either require values of higher-order parameters that appear in the asymptotic mean squared error formula, which are estimated or fixed, (e.g., setting $\rho=-1$), or use bootstrap schemes to circumvent having to do so.

For the \cite{Hall.Welsh:1985} class defined in \cref{eq:HallWelshclass}, for example, choosing the intermediate sequence $n_u$ so that $\lambda < \infty$ and the asymptotic mean squared error is $\xi^2/n_u + A^2(n/n_u)/(1-\rho)^2$ , gives 
 \begin{align}
 n_u = \left\lfloor \frac{(1-\rho)^2 n^{-2\rho}}{-2\rho \beta^2}\right\rfloor^{{1}/({1-2\rho})}\label{eq:kopt}
 \end{align}
as the optimal number of threshold exceedances. This is found by inserting estimates of $\beta$ and $\rho$ --- though the estimation of these second-order quantities requires another choice of intermediate sequence. Algorithm~1 of \cite{Caeiro.Gomes:2016} provides a fixed-point iteration scheme to estimate $n_u$. Since risk measures are typically high quantiles, minimization of the asymptotic mean squared error for estimation of $\xi$ might appear to be inappropriate, but the method appears to be competitive in simulations.

\cite{Dupuis.Victoria-Feser:2003} propose selecting the threshold by choosing the number of exceedances $n_u$ that minimizes a weighted prediction error criterion for Pareto tails, and in \citet{Dupuis.Victoria-Feser:2006} couple this with robust estimation of the shape parameter, which yield threshold-dependent weights for the observations. We did not implement this latter proposal.

\subsection{Bootstrap methods}

\citet[][\S4]{Hall:1990} was the first to suggest the use of resampling to evaluate the asymptotic mean squared error of Hill's estimator and choose $n_u$. 
Taking $m<n$ observations and $n_0$ initial exceedances, 
the bootstrap algorithm is as follows:
\begin{enumerate}
 \item estimate the shape parameter using Hill's estimator with the $n_0$ largest order statistics;
\item perform $B$ bootstrap replications as follows: resample $m =\mathrm{o}( n)$ observations with replacement, compute Hill's estimator for the $n_m=1, \ldots, m$ largest order statistics, and denote the shifted estimates for the $b$th replicate by $d^{(b)}_{n_m}=(\hat{H}^{(b)}_{m, n_m} - \hat{H}_{n, n_0})^2$;
\item for each value of $n_m$, average the $d^{(b)}_{n_m}$ over all bootstrap replications, and select the value $\hat{n}_m$ that minimizes the mean squared error.
\item For given $\rho$, compute the optimal number of exceedances for the full sample, i.e., $\hat{n}_u = \hat{n}_m(n/m)^{-2\rho/(1-2\rho)}$.
\end{enumerate}
The method is sensitive to the choices of $n_0$ and $m$, which are left to the user \citep{Caeiro.Gomes:2014}. Our simulations use the default values of $n_0 = 2n^{1/2}$ and $m=\lfloor n^{0.955}\rfloor$.

The bootstrap procedure requires the index of second-order regular variation to be known or estimated; \cite{Hall:1990} took $\rho=-1$. Fixing $\rho$ can be avoided by using a double bootstrap \citep{Draisma:1999}, which requires subsamples of sizes $m_1$ and $m_2 = n(m_1/n)^\omega < m_1$ with $\omega > 1$. If $\omega=2$, then $m_2=m_1^2/n$ and $\hat{n}_u^2(m_1)/\hat{n}_u(m_2) = \hat{n}_u(n) \{1+\mathrm{o}_p(1)\}$ as $n \to \infty$ \citep{Gomes.Oliveira:2001}. Ostensibly optimal values of $\hat{n}_u(n)$ may turn out to be larger than $n$ or may equal zero. While some authors suggest independent bootstrap runs, picking $m_2$ draws from the $m_1$ samples gives better precision, which might be further improved by bootstrap recyling \citep{Ventura:2002}.
 In summary, the optimal choice of $n_u(n)$ is based on the asymptotic relation between the latter and $\hat{n}_u(m_1)$ and $\hat{n}_u(m_2)$; the relationship also yields an estimate of $\rho$ as a by-product.

Other semiparametric shape estimators may have bias properties similar to those of Hill's estimator \citep[cf.][and references therein]{Caeiro.Gomes:2014}. \cite{Draisma:1999} proposed using the double bootstrap for the moment estimator of \cite{Dekkers.deHaan:1989}, whereas \cite{Danielsson:2001} considered the difference between de Vries' $\hat{\xi}^{\mathrm{v}}$ \citep{deHaan.Peng:1998} and Hill's estimators. Other choices of intermediate statistics are considered in \cite{Gomes.Oliveira:2001}, who compare the performance of different bootstrap methods for semiparametric estimation by simulation.

Estimation of second-order parameters can add appreciable variability, so some authors have followed \cite{Hall:1990} and fixed $\rho$. \cite{Schneider:2021} proposed the SAMSEE estimator of $n_u$, obtained by combining the generalized jackknife estimator $2\hat{\xi}^{\mathrm{v}}_{n,n_u}-H_{n,n_u}$ to estimate $\xi$ and a bias estimator obtained from the difference between smoothed Hill estimators over different windows; the result estimates the asymptotic mean squared error for $\rho=-1$. They report that their threshold selection procedure is competitive, with stable performance over a range of sample sizes and heavy-tailed distributions.

\cite{Drees.Kaufmann:1998} base threshold selection on calculations of the convergence rate of the maximum random fluctuation of the Hill estimator, $\max_{i} i^{1/2} | H_{n,i} - \xi - b_{n,i}| = \mathrm{O}_p\{\log^{1/2}(\log n)\}$ for the bias $b_{n,i}$, but \cite{Caeiro.Gomes:2016} claim that this is sensitive to tuning parameters; in our simulations, it usually failed to return results whe using the default values for them, and its performance seems to be subpar.

\subsection{Exponential approximations}

Under the assumption of regular variation, and with $\xi>0$, the log-spacings form an approximate exponential sample, so \cite{Hill:1975} suggested increasing the number of order statistics until the exponential model is found to be inadequate. As the departure from the theoretical model is very gradual, this typically selects $n_u$ to be too large \citep{Hall.Welsh:1985}.

\cite{Guillou.Hall:2001} also consider approximating the distribution of Hill's estimator and a modification thereof based on linear combination of log spacings, obtaining a sum of exponential variables and a bias term to be minimized. The sum of squares of the zero-mean statistic is computed over a moving window and the chosen threshold is the smallest $n_u$ for which the estimate exceeds a critical value.

\cite{Beirlant.Vynckier.Teugels:1996a} note that, when the sample exceedances are tail-equivalent to a Pareto distribution, the shape parameter $\xi>0$ can be viewed as the slope in an exponential quantile-quantile plot with plotting positions $-\log\{j/(n_u+1)\}$ against order statistics $Y_{(n-j)}$, which leads to the Hill estimator and kernel-weighted variants \citep{Csorgo.Deheuvels.Mason:1985}. Under second-order regular variation with $\rho \leq 0$, they propose finding the value $n_u$ that minimizes
\begin{align}
 \mathsf{MSE}(H_{n,n_u}) = \frac{1}{n_u} \sum_{j=1}^{n_u} w_{j,n_u}^{\mathrm{opt}} \left\{\log\left(\frac{Y_{(n-j+1)}}{Y_{(n-n_u)}}\right) - H_{n,n_u}\log \left( \frac{n_u+1}{j}\right)\right\}^2. \label{eq:mseBeirlant}
\end{align}
The weights $w_{j,n_u}^{\mathrm{opt}}$ depend on $\rho$ and provide an estimate of the asymptotic mean squared error of the Hill estimator. \cite{Beirlant.Vynckier.Teugels:1996a} propose an iterative procedure with an initial estimate of $n_u$ obtained from an unweighted version of \cref{eq:mseBeirlant}, estimate $\rho$ from $m>n_u$ order statistics, and then alternate between estimating $n_u$ and $\rho$. This procedure is generalized to real-valued $\xi$ in \cite{Beirlant.Vynckier.Teugels:1996b} using the generalized quantile regression estimator. The weights depend on $\xi$ when it is negative, and the estimator of $\rho$ is erratic and may return positive values. Both $m$ and $\rho$ must be constrained to avoid nonsense output, and, as for any second-order parameter, unrealistically large samples are needed for reliable estimation of $\rho$. Moreover the optimal weights can be negative, leading to invalid mean squared errors \citep{Gomes.Oliveira:2001}. Sample paths of \cref{eq:mseBeirlant} are variable, and the procedure does not account for the large uncertainty of the MSE estimates.

Still under an assumed Pareto upper tail, \cite{Beirlant:1999} propose a nonlinear exponential regression procedure to estimate the scaling sequence, shape and second-order regular variation parameters based on the asymptotic representation of log-spacings of upper order statistics as independent exponential variables \citep{Beirlant:1999},
\begin{align*}
 Z_j = j(\log Y_{(n-j+1)} - \log Y_{(n-j)}) \stackrel{\cdot}{\sim} \mathsf{exp}\left[\xi + b_{n,n_u} \{j/(n_u+1)\}^{-\rho}\right], \qquad j=1,\ldots, n_u.
\end{align*}
A similar model is considered in \cite{Feuerverger.Hall:1999}. The parameters $\xi$, $\rho$ and $b_{n,n_u}$ can be estimated by maximum likelihood subject to the constraints $\xi>0$, $\rho < 0$ and $b_{n,n_u}>0$, perhaps with further constraints on $\rho$ to reduce bias and avoid nonsensical values. \cite{Beirlant:2002} consider nonlinear least squares estimators under additive noise by joint estimation and by a two-stage procedure obtained by substituting a consistent estimator of $\rho$ based on $n_0$ order statistics considering the least-square bias with $\rho=-1$, then estimating other parameters by maximum likelihood. Under the Hall--Welsh class of distributions~\eqref{eq:HallWelshclass}, they derive the optimal $n_u$ as a function of these parameters and scale parameter $b$, shape $\xi$ and second-order regular variation index $\rho$. They suggest an ad hoc automation of their approach by looking at the median estimator obtained by applying the formula for $n_0=3, \ldots, n/2$.

\cite{Goegebeur.Beirlant.deWet:2008} use kernel weighting to derive alternative tail index estimators and use these to approximate the asymptotic mean squared error of the Hill estimator, with a penalty term. \cite{Bladt:2020} use the Rényi representation to define lower-trimmed Hill estimators, which include Hill's estimator as a special case. Focusing on Hall--Welsh distributions, they derive an explicit relationship between the optimal number of exceedances for the Hill and the lower trimmed Hill estimators under second-order regular variation, and recommend taking the number of exceedances that minimizes the average left-trimmed estimator variance, which is smooth and depends less on $n_u$ than does the Hill estimator. They then use the explicit relationship between estimators to find the corresponding $n_u$ for Hill's estimator, assuming $\rho=-1$. The method crucially depends on the values for $n_u$ selected for the left-trimmed estimator, which they take to be at least $n/5$. Threshold selection is followed by check of Pareto tails using ratio statistics.

\subsection{Alternative procedures}

The sample paths of the Hill estimator as a function of $n_u$ are non-differentiable, whereas those of the random block maximum estimator proposed by \cite{Wager:2014} are infinitely differentiable, thus facilitating visual inspection and threshold selection. The estimator of \cite{Wager:2014} is also asymptotically normal under second-order regular variation with $\rho < 0$. \citeauthor{Wager:2014} recommends using empirical risk minimization to choose $n_u$: the procedure is based on a finite-difference approximation of the squared derivative of the process, subject to a penalty term, which corresponds to the expected bias. Although the computational cost is higher than for Hill's estimator, it can be kept reasonable by restricting $n_u$.

\section{Numerical results} \label{sec:numresults}
\subsection{Simulation study}\label{sec:numerical}

Assessing the performance of threshold selection methods is hard because there is no right answer, but there are approaches to avoid. For example, data rarely show sharp changes in tail behaviour, so spliced distributions are typically unrealistic as a basis for comparing methods. Moreover, since the tail index $\xi$ is rarely of direct interest, it is unwise to focus on it, particularly on its limiting value, which can provide a poor fit in finite samples (\Cref{sec:penultimate}), while the asymptotic bias involving the second-order conditions would be unknown in practice. Risk measures of more practical interest are typically exceedance probabilities or high quantiles, whose sampling distributions are usually quite skewed \citep[cf.][]{Belzile.Davison:2022}. Stability of these estimates is crucial, as showcased in \Cref{figthstab}.

\Cref{sec:simulationstudy} describes the results of a very extensive simulation study, whose general findings we summarise here. Our study considers a wide range of distributions, and targets their 0.999 quantiles using various sample sizes. A key aspect is the sensitivity of threshold selection procedures to the distribution and sample size, and whether the chosen thresholds depend much, if at all, on the scenario.
The threshold selection methods we consider are either parametric (where parameters are estimated via maximum likelihood) or semiparametric (mostly based on Hill's estimator or variants thereof, coupled with Weissman's quantile estimator).

In the simulations for the parametric methods, we use a grid $\mathcal{U}$ at equispaced quantile levels, estimated empirically for each dataset to ensure constant numbers of exceedances. We devise an oracle that finds the threshold for which the 0.999 quantile estimator is closest to the true one. This returns thresholds that are usually between the 0.87 and the 0.9 quantiles, suggesting that the variability of the estimators is as important as threshold selection. The performance of the maximum likelihood estimator degrades  when $\xi \geq 0.5$, but such tails arise rarely in applications. Unless there are strong penultimate effects, most methods prioritize large samples, and many of the automated procedures return the lowest possible threshold, with the mode for most scenarios being the 0.8 quantile. The selection methods of \cite{Collings:2025} and the maximum posterior weight and Bayesian model averaging estimator of \cite{Northrop.Attalides.Jonathan:2017} perform reasonably well,  though no procedure is uniformly best across all distributions. Rounding and serial correlation have little impact on threshold selection, although autocorrelation leads to large bias in the return level estimates unless accounted for.

Some of the semiparametric threshold selection procedures are promising, but again there is no clear winner. Some procedures yield erratic results, fail to converge, and are very slow. We give advice on which method works best depending on the user case, and list some that should be disregarded altogether. Methods that fix the second-order regular variation parameter tend to be more robust, even if this induces misspecification and bias. Methods that return too small values of $n_u$ lead to overly variable results. Bootstrap estimation of the AMSE via bootstrap is expensive, but works seemingly well. The method of \cite{Guillou.Hall:2001} and the SAMSEE estimator of \cite{Schneider:2021} are also competitive.

As one would expect, increasing the sample size $n$ leads to both larger $n_u$ and smaller sample fractions kept for inference, i.e., $n_u/n$ decreases with $n$. The semiparametric methods are quite sensitive to the data passed to them, with many giving results that depend strongly on the choice of candidate thresholds. Restricting threshold selection from the set of order statistics to a range of say $n_u \in [20, 1000]$ can reduce both the variance of quantile estimates and generally the variance for most methods, although it biases shape parameter estimation when $\xi > 0.5$.

\subsection{Data application} \label{sec:data_application}

\Cref{tab-parametric-padova,tab-semiparametric-padova} report results for the application of most of the threshold selection procedures discussed here to the Padova rainfall data; results  for four other data sets are in \Cref{sec:dataapp}. Although a wide variety of thresholds and thus of shape parameters are selected, the semiparametric estimates of the shape based on Hill's estimator are generally higher than their maximum likelihood counterparts, but mostly give lower quantile estimates; the shape estimates are largely responsible for these differences, whatever the threshold. Many of the parametric methods return the lowest threshold tested: one could consider lower thresholds, but this would yield a poor fit to the largest observations.

 \begin{table}[ht!]
 \caption{Parametric threshold selection for the Padova rainfall data, using candidate thresholds $u$ equispaced on the quantile levels ranging 0.8 to 0.995 in increments of 0.005. The 100-year return levels (in mm) and shape parameters returned are maximum likelihood estimates, except for the Bayesian model averaging of \cite{Northrop.Attalides.Jonathan:2017} which gives posterior means based on validation weights at the 0.99 quantile.}
\centering

\begin{tabular}{lccrcc}
\toprule
method & $u$ & level (\%) & $n_u$ & shape & return level\\
\midrule
\cite{Caers.Beirlant.Maes:1999} & 13.80 & 80.5 & 645 & 0.07 & 108.4\\
\cite{Gonzalo.Olmo:2004} & 53.09 & 99.0 & 34 & 0.09 & 117.4\\
\cite{Thompson:2009} & 18.50 & 86.5 & 446 & 0.09 & 110.6\\
\cite{Suveges.Davison:2010} & 16.00 & 83.5 & 544 & 0.08 & 109.8\\
\cite{Northrop.Coleman:2014} & 53.09 & 99.0 & 34 & 0.09 & 117.4\\
\cite{Langousis:2016} & 13.58 & 80.0 & 663 & 0.08 & 109.5\\
\cite{Wadsworth:2016} & 13.58 & 80.0 & 663 & 0.08 & 109.5\\
\cite{Northrop.Attalides.Jonathan:2017} &  &  & 368 & 0.13 & 119.0\\
\cite{SilvaLomba.FragaAlves:2020} & 39.97 & 97.5 & 83 & 0.21 & 122.2\\
\cite{Kiran.Srinivas:2021} & 37.93 & 97.0 & 100 & 0.24 & 124.6\\
\cite{Varty.Tawn:2021} & 13.58 & 80.0 & 663 & 0.08 & 109.5\\
\cite{Gamet.Jalbert:2022} (beta EGP model) & 21.60 & 89.0 & 360 & 0.13 & 115.2\\
\cite{Murphy.Tawn.Varty:2024} & 13.58 & 80.0 & 663 & 0.08 & 109.5\\
\cite{Collings:2025} & 22.49 & 90.0 & 332 & 0.12 & 114.8\\
\bottomrule
\end{tabular}
\label{tab-parametric-padova}
\end{table}

\begin{table}[ht!]
 \caption{Semiparametric threshold selection for the Padova rainfall data. Columns give the selected threshold $u$, the number of exceedances $n_u$, the shape parameter based on Hill's estimator (except for \cite{Wager:2014}), and the estimated 100-year return levels (in mm) computed using Weissman's estimator.}
 \centering
 
\begin{tabular}{lcrcc}
\toprule
method & $u$ & $n_u$ & shape & return level\\
\midrule
\cite{Hall.Welsh:1985} & 48.5 & 41 & 0.31 & 139.9\\
\cite{Hall:1990} & 33.8 & 131 & 0.32 & 146.7\\
\cite{Beirlant.Vynckier.Teugels:1996b} & 13.5 & 665 & 0.21 & 48.0\\
\cite{Drees.Kaufmann:1998} & 117.6 & 1 & 0.29 & 107.2\\
\cite{Guillou.Hall:2001} & 33.4 & 138 & 0.32 & 144.2\\
\cite{Danielsson:2001} & 58.2 & 28 & 0.24 & 118.1\\
\cite{Dupuis.Victoria-Feser:2003} (non-robust) & 76.4 & 10 & 0.20 & 113.7\\
\cite{Reiss.Thomas:2007} ($l_1$) & 24.2 & 297 & 0.37 & 180.1\\
\cite{Reiss.Thomas:2007} ($l_2$) & 16.0 & 548 & 0.51 & 347.8\\
\cite{Goegebeur.Beirlant.deWet:2008} & 30.2 & 179 & 0.34 & 159.7\\
\cite{Clauset.Shalizi.Newman:2009} & 31.6 & 167 & 0.32 & 146.0\\
\cite{Gomes.Figueiredo.Neves:2012} & 48.0 & 45 & 0.30 & 136.1\\
\cite{Gomes:2013} (sample paths) & 48.8 & 39 & 0.32 & 143.1\\
\cite{Caeiro.Gomes:2014} & 25.4 & 265 & 0.37 & 175.7\\
\cite{Wager:2014} & 16.1 & 541 & 0.25 & 69.8\\
\cite{Caeiro.Gomes:2016} (AMSE) & 36.0 & 114 & 0.31 & 139.2\\
\cite{Danielsson:2019} (eye-balling) & 78.8 & 7 & 0.25 & 118.3\\
\cite{Danielsson:2019} (MAD) & 36.8 & 107 & 0.30 & 136.2\\
\cite{Danielsson:2019} (KS) & 39.3 & 88 & 0.30 & 134.0\\
\cite{Bladt:2020} & 76.6 & 10 & 0.20 & 114.0\\
\cite{Schneider:2021} (SAMSEE) & 43.5 & 62 & 0.29 & 131.9\\
\bottomrule
\end{tabular}

 \label{tab-semiparametric-padova}
 \end{table}

\section{Discussion} \label{sec:discussion}

\subsection{Practical considerations}
\subsubsection{Multiple hypothesis testing}


The starting point is typically a predefined sequence of potential ordered thresholds $\calU = \{u_1,\ldots, u_J\}$, often high sample quantiles. The highest is chosen to be low enough that estimation based on its exceedances is reliable, and the lowest often lies well inside the dataset.
In either case threshold selection is implicitly seen as equivalent to a sequence of tests of the hypotheses $\mathscr{H}_j: Y-u_j \mid Y>u_j \sim \ddistGP(\sigma_j, \xi)$, for $u_j\in\calU$. The threshold-stability of the GPD implies that the hypotheses are nested, i.e., $\mathscr{H}_J \subset \cdots \subset \mathscr{H}_1$. If $\mathscr{H}_j$ is true, then $\mathscr{H}_l$ is true for all $l\geq j$, so we seek the smallest $k$ for which $\mathscr{H}_k$ is true. Since $\mathscr{H}_1,\ldots, \mathscr{H}_{k-1}$ are then false, it seems natural to deal with the multiple testing using a procedure such as ForwardStop or StrongStop \citep{GSell:2016}, though these rely on independence of the null $p$-values. Under this assumption both procedures control the false discovery rate, and StrongStop also controls the family-wise error rate. In threshold selection this error rate is the probability that the selected threshold is too high, and the overall power is related to the probability that it is too low. The procedures described above do not provide independent null $p$-values, so the proofs of their properties fail.

Since the $\mathscr{H}_j$ specify the models only above the $u_j$, the samples differ from one model to the next and so likelihood ratio tests or information criteria cannot be applied. It would be possible to use only exceedances above $u_J$ as test samples, though the asymptotics would be non-standard. Extended models such as those described in \Cref{sec:splicing}, which specify the fit over the whole domain or at least above $u_1$, are not subject to these restrictions. These models are generally difficult to fit but have the advantage that the threshold is amenable to selection using information criteria.

\subsubsection{Many series}

The number of separate variables or time series for which thresholds must be chosen is large in financial data, in spatial settings, and in other large-scale applications. Such settings require some degree of automation, in particular since individual inspection of graphs for huge numbers of series is precluded. Another aspect, typically ignored, is any dependence between series, which should lead one to select similar thresholds. Smoothing of some sort could improve estimation, and joint tests of adequacy could be constructed, and perhaps used to borrow strength for threshold selection.

\subsubsection{Nonstationarity}

In many cases, the marginal distribution varies over time or depends on covariates. In such cases applied workers often attempt to obtain approximately stationary data by analysing time windows, such as seasons, and analysing these separately.

An alternative is a non-stationary threshold \citep[e.g.,][Chapters~6--7]{Coles:2001}. This can lead to difficulties in obtaining marginal or conditional return levels below the threshold, but there are many such proposals in the literature \citep[e.g.,][]{Coelho:2008,Northrop.Jonathan:2011}, including the use of quantile regression using an asymmetric Laplace distribution \citep{Youngman:2019,Fasiolo:2021} or extremal random forests \citep{Gnecco:2024}. These methods fix the threshold level relatively low (0.8 or 0.9, say), as such models require appreciable data for reliable estimation.

\subsubsection{Ties}

Much environmental data is recorded with limited accuracy, which prevents use of estimators based on log differences of order statistics. This is awkward in semiparametric settings, but can be handled in a likelihood framework by treating observations as interval-censored; for an example of these impacts and mitigation strategies, see \cite{Varty.Tawn:2021}.

\subsection{General}

Our review highlights the inherent limitations of threshold selection procedures. No approach is entirely satisfactory, but the simulation studies and implementations reveal some overarching features that are worth highlighting.

While asymptotic normality of Hill-type estimators is obtained under an assumption of second-order regular variation, estimates of the second-order parameter $\rho$ are typically far too noisy to be useful: more stable methods are obtained by fixing it to a negative value, or bypassing its estimation completely, as in \cite{Wager:2014}. Many algorithms consider all potential choices of threshold, irrespective of the minimum number of exceedances needed for reliable estimation, or of the fact that only the largest observations should be selected.

Parametric methods are generally less easy to automate, and automatic procedures such as that of \cite{Langousis:2016} often give unsatisfactory results. We have made pragmatic choices in doing tests sequentially (from the highest selected thresholds) to reflect what seems to be common practice when interpreting threshold stability plots and related procedures. Some of the more promising methods are based on measures of fit and certain apparently \textit{ad hoc} methods \citep[e.g.,][]{Thompson:2009} seem to perform well in practice.

Are we barking up the wrong tree? In most cases the transition from the bulk of the distribution to the limiting tail behaviour is gradual, so there is no ``correct threshold'' to be chosen. This has led to the emergence of sub-asymptotic, extended generalised Pareto models, but it is unclear whether the associated increase in uncertainty for the main quantities of interest is a price worth paying.

\section*{Acknowledgements}
We thank Sonia Alouini for valuable discussions and for detailed feedback on an earlier draft, and Marco Marani for providing the Padova data. The work was supported by the Swiss National Science Foundation and by the Natural Sciences and Engineering Research Council through grants RGPIN-2022-05001 and DGECR-2022-00461.

\section*{Reproducibility}
This paper used code from \textbf{R} packages \href{https://cran.r-project.org/web/packages/mev/index.html}{\texttt{mev}} \citep{mev}, \href{https://cran.r-project.org/web/packages/tea/index.html}{\texttt{tea}} \citep{tea} and \href{https://cran.r-project.org/web/packages/threshr/index.html}{\texttt{threshr}} \citep{threshr}. Code for reproducing the results of this paper is available at \href{https://github.com/lbelzile/choosing-threshold}{\texttt{https://github.com/lbelzile/choosing-threshold}}.
\bibliographystyle{imsart-nameyear}
\bibliography{threshold}

\clearpage
\begin{appendix}
\section{Exploratory analysis of the Padova data}
\begin{figure}[b!]
 \centering
 \includegraphics[width = 0.9\textwidth]{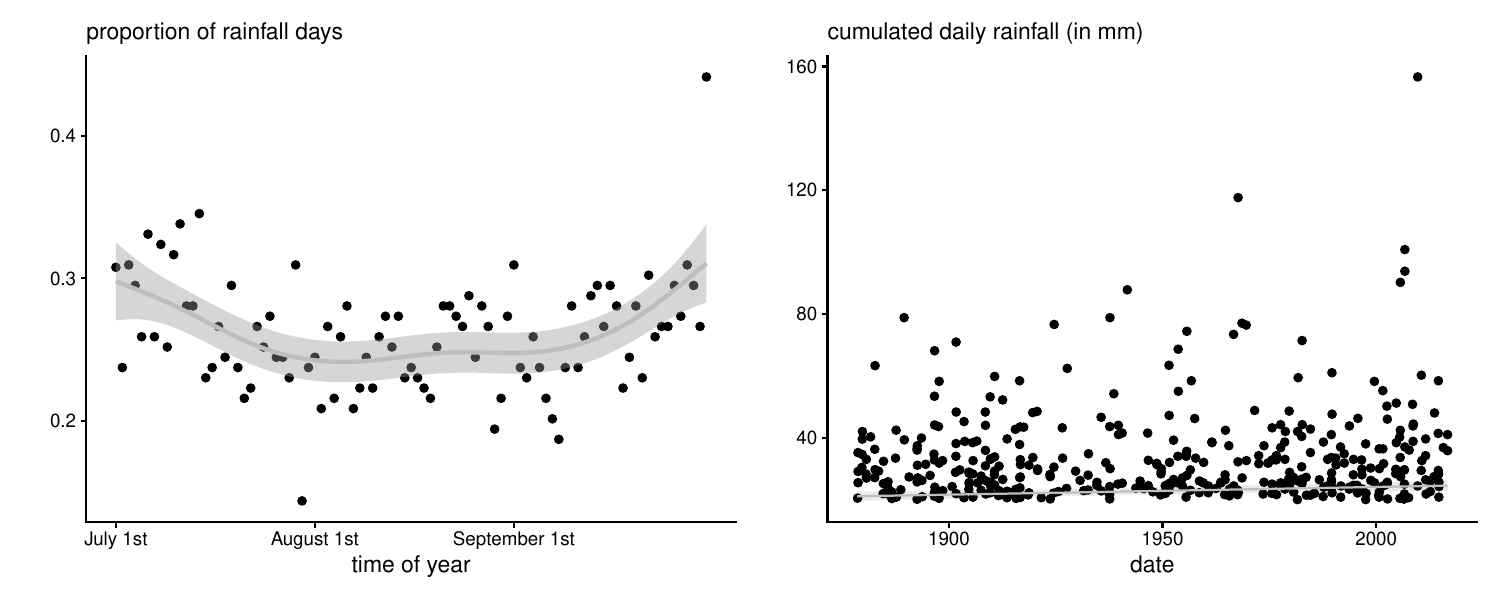}
 \caption{Analysis of seasonality for the Padova data, using only the period for July--September. The left panel shows the proportion of rainy days as a function of the day of the year, with a generalized additive model smooth. The right panel shows the exceedances above 20mm of rainfall, and the estimated quantile regression curve with pointwise Wald 95\% confidence intervals, for the 90\% of non-zero rainfall.}
 \label{edapadova}
\end{figure}

\Cref{edapadova} shows the seasonal and temporal trends for the Padova data. Quantile regression indicates some residual seasonality and nonstationarity; in particular there is a small but statistically significant increase of around 3.4~mm in the 90\% quantile from 1950 onwards, but this is smaller than the seasonal variability for the period.

\section{Penultimate approximation} \label{sec:penultimate}

\Cref{sec:secondorder} gives a brief account of the theoretical conditions under which the GPD provides a limiting distribution for exceedances over high thresholds, and mentions that under more restrictive but nevertheless reasonable conditions, the GPD also provides a penultimate approximation, albeit with a shape parameter that does not equal its limiting value. Such results are useful, but it is important to appreciate that empirical fitting can improve on both the limiting and penultimate approximations. We illustrate this by maximum likelihood fitting of the GPD to samples of $N=10^7$ Gaussian observations lying above the $q=0.9$, 0.95 and 0.99 quantiles, $u_q$, say, of the Gaussian distribution. The idea is not to reproduce sample sizes that might be met in practice, but to assess how closely the generalized Pareto model can match the `true' truncated Gaussian distribution. The quality of the resulting approximation depends on $N$; in principle the fitted shape parameter $\xi_N$ must tend to zero as $N\to\infty$, but in practice the tail of the truncated distribution is so short that $\xi_N$ is appreciably negative even with this huge value of $N$.

\begin{figure}[t!]
 \centering
 \includegraphics[width=\textwidth]{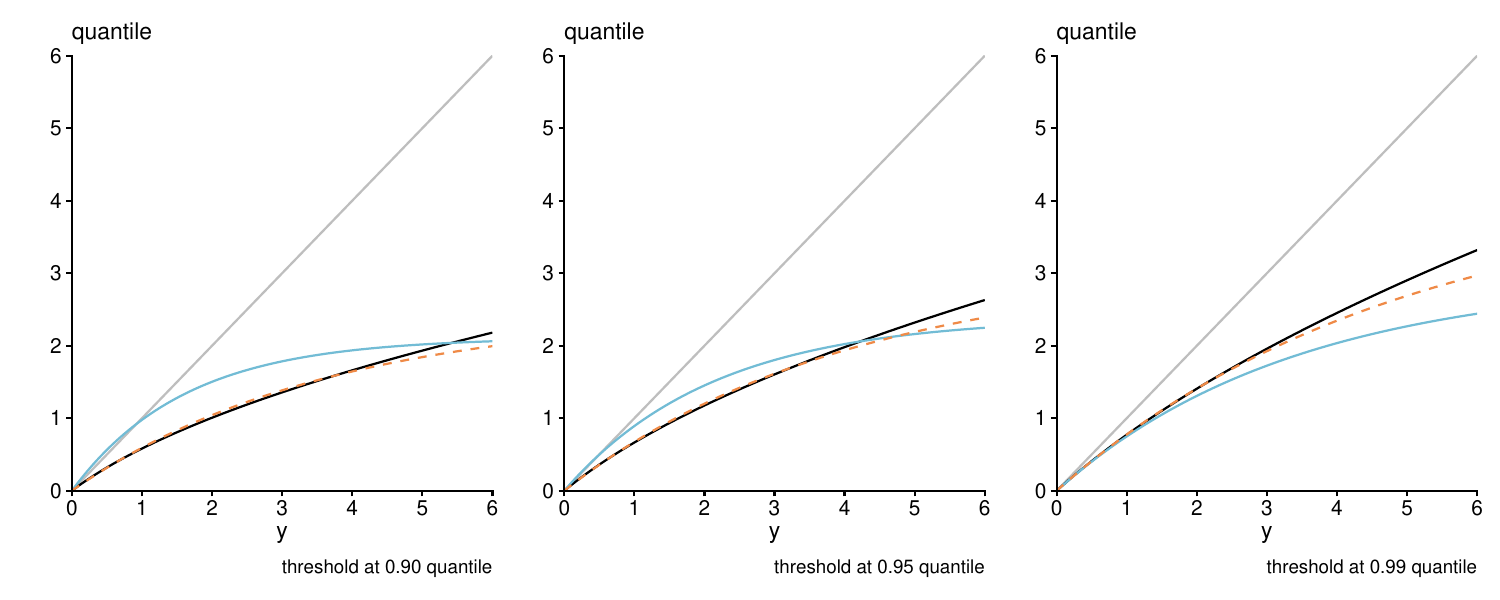}
 \caption{Penultimate fits for ten million exceedances for the Gaussian model with thresholds at its 0.9, 0.95 and 0.99 quantiles (left to right). The limiting distribution is scaled so that the exceedance probability for $y$ is $10^{-y}$. The panels show scaled excess quantile plots for the true truncated normal (black), penultimate approximation (full blue), best fitted generalized Pareto (dashed orange) and limiting exponential (grey straight line with unit slope). Curves below the diagonal correspond to the implied excess quantiles from generalized Pareto or the true model. The fitted shape parameters $\hat\xi$ are $-0.123$, $-0.117$ and $-0.088$, while the penultimate values $\xi_t$ are $-0.270$, $-0.203$ and $-0.127$.}
 \label{fig:penultimate_effects}
\end{figure}

\Cref{fig:penultimate_effects} shows how well the quantiles of the left-truncated Gaussian distributions are matched by those of the limiting exponential distribution, the penultimate approximation with shape parameter $\xi_t$, and the generalized Pareto distribution fitted by maximum likelihood estimation based on the $N$ exceedances.
Although negative, the fitted shape parameters $\xi_N$ are appreciably closer to zero than the penultimate values, and the corresponding quantiles track the true quantiles remarkably well for tail probabilities down to $10^{-4}$ or so. Maximum likelihood estimators based on samples of size $n_{u_q}\ll N$ would be expected to have means roughly $\xi_N$, rather than the limiting or penultimate values of $\xi$, with $O(n_{u_q}^{-1})$ bias.


%
%

Similar computations (large-sample maximum likelihood estimates and penultimate approximations) for the shape parameter of a series of distributions (cf. \Cref{sec:targetdist}) considered in the simulation study are reported in \Cref{tab_penult_shape}. They show the same phenomenon, namely that the maximum likelihood values are closer to the limit than are the penultimate values. 

%
%
%

\section{Simulation study} \label{sec:simulationstudy}
In this section we present results from an extensive simulation study that covers some scenarios encountered in applications, including data recorded with limited precision (rounding), and dependent time series with clusters. The data are simulated from a variety of parametric models with different shape parameters and rates of convergence to the  limiting model.

We consider return levels for quantile level $q=0.999$, as this could be reasonably inferred from the data while being extreme enough that empirical estimation is infeasible or too noisy to be useful.  
As some of the distributions considered have infinite first or second moments, or very different scales, we consider relative errors.

We compare performance in terms of bias and variance  (through sampling distribution and absolute bias measures), all relative to the true value. When relevant, we also report the computing time. We also assess the variability of the proportion of exceedances retained or the number of order statistics above $u$ kept by the selection procedure. Some methods may fail to return a threshold, and the percentage of such failures impacts both their practical usefulness and the ease with which they can be studied by simulation.

\subsection{Target distributions} \label{sec:targetdist}

Simulation studies should take distributions whose tail behaviour is also plausible in applications --- for example, heavy-tailed cases with $\xi \approx 1$ are unrealistic outside of specific financial or network applications, whereas most environmental phenomena (e.g., temperature, wind speed, rainfall) would have reasonable values of $\xi \in [-0.5, 0.25]$ in practice. Many papers on threshold selection have used mixture models: for example, \cite{Lee:2015} considered non-overlapping distributions with a generalized Pareto tail above $u$, and another distribution, right-truncated at $u$, for the bulk. Such models have a sharp discontinuity at the threshold $u$, a feature which simplifies threshold selection but is  seldom observed in practice.
We therefore instead consider distributions featured in simulation studies of \cite{Choulakian.Stephens:2001} (a--d), \cite{Schneider:2021} (i--m), and extended generalized Pareto models with exact generalized Pareto limits (e, n--p).  Here is our full list: 
\begin{multicols}{2}
\begin{enumerate}[label={(\alph*)}]
 \item gamma with shape $2$ and scale $1$;
 \item standard lognormal;
 \item Weibull with scale $1$ and shape $0.75$;
 \item Weibull with scale $1$ and shape $1.25$;
 \item piecewise generalized Pareto \citep{Northrop.Coleman:2014} with shape $0.25$ up to $u=1.25$ and $-0.25$ above
 \item standard exponential ($\xi=0$)
\item generalized Pareto with shape $\xi=0.15$
\item generalized Pareto with shape $\xi=-0.15$
\item folded Student-$t$ with 6 degrees of freedom $|T_{6}|$ ($\xi=1/6$);
 \item standard Fréchet with shape $\xi=0.5$;
 \item standard Cauchy ($\xi=1$) truncated on $\mathbb{R}^{+}$;
 \item loggamma with density function $\log(x) x^{-2}$ for $x \geq 1$ ($\xi=1$);
 \item Burr with survival function $(1+x^{2})^{-1}$ for $x>0$ ($\xi=0.5$);
\item piecewise generalized Pareto \citep{Northrop.Coleman:2014} with shape $-0.5$ up to the $0.9$ quantile and $0.25$ above;
\item third extended generalized Pareto model of \cite{Papastathopoulos.Tawn:2013} (power model) with $\xi=-0.2$, unit scale and $\kappa = 0.25$;
\item exponential tilting extended generalized Pareto model with shape $\xi =0.2$ and $\kappa=0.1$.
\end{enumerate}
\end{multicols}

 The distributions (f--h) are threshold-stable, while the penultimate models (n--p) converge rather slowly to the generalized Pareto limit. \Cref{tab_penult_shape} shows that the average maximum likelihood estimates for the shape for varying quantile levels vary very little, and can differ markedly from the limiting shape value and the penultimate approximation, though the latter tends to be closer to the best fitting model given by the \textsc{mle}. The fact that the latter barely changes suggests that methods ought to take lowest threshold in most cases, though this does not take into account goodness-of-fit considerations, and small-sample bias is more acute when using higher thresholds. \Cref{tab_penult_quantratio} shows that, unless convergence to the limit is quite slow (cf. distribution l), the approximation for the 0.999 quantile is quite accurate, that the penultimate approximation can be markedly better than using the limit shape, but that the penultimate approximation is not a panacea, especially at quantiles 0.8 and 0.9.

%
\begin{table}[ht!]
\centering
\caption{Shape parameters for varying quantile levels (0.8, 0.9, 0.95, 0.99) obtained using \cite{Smith:1987} penultimate approximation, maximum likelihood estimation with samples of 1M exceedances, and the limiting value of $\xi$ for models without exact generalized Pareto tails.} \label{tab_penult_shape}
\begin{tblr}[ 
] 
{ 
colspec={Q[c]Q[r]Q[r]Q[r]Q[r]Q[r]Q[r]Q[r]Q[r]Q[r]},
hline{2}={3-4,7-8}{solid, black, 0.05em},
hline{3}={1-10}{solid, black, 0.05em},
hline{2}={2,6,10}{solid, black, 0.05em, l=-0.5},
hline{2}={5,9}{solid, black, 0.05em, r=-0.5},
hline{1}={1-10}{solid, black, 0.1em},
hline{14}={1-10}{solid, black, 0.1em},
cell{1}{1}={}{halign=c},
cell{1}{10}={}{halign=c},
cell{1}{2}={c=4}{halign=c},
cell{1}{3}={}{halign=c},
cell{1}{4}={}{halign=c},
cell{1}{5}={}{halign=c},
cell{1}{6}={c=4}{halign=c},
cell{1}{7}={}{halign=c},
cell{1}{8}={}{halign=c},
cell{1}{9}={}{halign=c},
} 
& penultimate & & & & mle & & & & limit \\
model & $80\%$ & $90\%$ & $95\%$ & $99\%$ & $80\%$ & $90\%$ & $95\%$ & $99\%$ & $ $ \\
a & $-0.11$ & $-0.07$ & $-0.04$ & $-0.02$ & $-0.04$ & $-0.03$ & $-0.02$ & $-0.01$ & $0\hphantom{.00}$ \\
b & $0.32$ & $0.3\hphantom{0}$ & $0.28$ & $0.25$ & $0.28$ & $0.26$ & $0.25$ & $0.22$ & $0\hphantom{.00}$ \\
c & $0.21$ & $0.14$ & $0.11$ & $0.07$ & $0.11$ & $0.09$ & $0.07$ & $0.05$ & $0\hphantom{.00}$ \\
d & $-0.12$ & $-0.09$ & $-0.07$ & $-0.04$ & $-0.06$ & $-0.05$ & $-0.04$ & $-0.03$ & $0\hphantom{.00}$ \\
i & $-0.08$ & $0\hphantom{.00}$ & $0.05$ & $0.11$ & $0.06$ & $0.09$ & $0.11$ & $0.14$ & $0.17$ \\
j & $0.43$ & $0.47$ & $0.49$ & $0.5\hphantom{0}$ & $0.48$ & $0.49$ & $0.5\hphantom{0}$ & $0.5\hphantom{0}$ & $0.5\hphantom{0}$ \\
k & $0.93$ & $0.98$ & $1\hphantom{.00}$ & $1\hphantom{.00}$ & $0.99$ & $1\hphantom{.00}$ & $1\hphantom{.00}$ & $1\hphantom{.00}$ & $1\hphantom{.00}$ \\
l & $1.22$ & $1.19$ & $1.17$ & $1.13$ & $1.17$ & $1.15$ & $1.13$ & $1.11$ & $1\hphantom{.00}$ \\
m & $0.38$ & $0.44$ & $0.47$ & $0.49$ & $0.47$ & $0.48$ & $0.49$ & $0.5\hphantom{0}$ & $0.5\hphantom{0}$ \\
o & $0.31$ & $0.01$ & $-0.1\hphantom{0}$ & $-0.18$ & $-0.1$ & $-0.15$ & $-0.18$ & $-0.2\hphantom{0}$ & $-0.25$ \\
p & $0.39$ & $0.36$ & $0.31$ & $0.23$ & $0.3\hphantom{0}$ & $0.27$ & $0.24$ & $0.21$ & $0.2\hphantom{0}$ \\
\end{tblr}

\end{table}

\begin{table}[ht!]
\centering
\caption{Ratio of estimated 0.999 quantile to the true quantile (in \%) for varying quantile levels (0.8, 0.9, 0.95, 0.99) obtained using \cite{Smith:1987} penultimate approximation, maximum likelihood estimation with samples of 1M exceedances, and the limiting value of $\xi$.} \label{tab_penult_quantratio}
{\scalebox{0.9}{
\begin{tblr}[ 
] 
{ 
colspec={Q[r]Q[r]Q[r]Q[r]Q[r]Q[r]Q[r]Q[r]Q[r]Q[r]Q[r]Q[r]Q[r]},
hline{2}={3-4,7-8,11-13}{solid, black, 0.05em},
hline{3}={1-13}{solid, black, 0.05em},
hline{2}={2,6,10}{solid, black, 0.05em, l=-0.5},
hline{2}={5,9}{solid, black, 0.05em, r=-0.5},
hline{1}={1-13}{solid, black, 0.1em},
hline{14}={1-13}{solid, black, 0.1em},
cell{1}{1}={}{halign=c},
cell{1}{10}={c=4}{halign=c},
cell{1}{11}={}{halign=c},
cell{1}{12}={}{halign=c},
cell{1}{13}={}{halign=c},
cell{1}{2}={c=4}{halign=c},
cell{1}{3}={}{halign=c},
cell{1}{4}={}{halign=c},
cell{1}{5}={}{halign=c},
cell{1}{6}={c=4}{halign=c},
cell{1}{7}={}{halign=c},
cell{1}{8}={}{halign=c},
cell{1}{9}={}{halign=c},
} 
& penultimate & & & & mle & & & & limit & & & \\
model & 80\% & 90\% & 95\% & 99\% & 80\% & 90\% & 95\% & 99\% & 80\% & 90\% & 95\% & 99\% \\
a & $90.2$ & $96.2$ & $98.5$ & $99.9$ & $99.5$ & $99.9$ & $100\hphantom{.0}$ & $100\hphantom{.0}$ & $109\hphantom{.0}$ & $104.8$ & $102.7$ & $100.6$ \\
b & $113.9$ & $109.2$ & $105.2$ & $100.9$ & $102.8$ & $101\hphantom{.0}$ & $100.1$ & $99.7$ & $50.5$ & $59.4$ & $68.3$ & $86.8$ \\
c & $128.8$ & $110.7$ & $104.4$ & $100.5$ & $102.2$ & $100.5$ & $99.9$ & $99.9$ & $77.3$ & $84.7$ & $90\hphantom{.0}$ & $97.1$ \\
d & $91.3$ & $96.3$ & $98.4$ & $99.8$ & $99.5$ & $100\hphantom{.0}$ & $100.1$ & $100.1$ & $113.3$ & $108\hphantom{.0}$ & $104.8$ & $101.2$ \\
i & $77.9$ & $88.8$ & $94.7$ & $99.4$ & $97.8$ & $99.5$ & $100.1$ & $100.2$ & $129.6$ & $116.5$ & $109.2$ & $102\hphantom{.0}$ \\
j & $83.3$ & $94.2$ & $98\hphantom{.0}$ & $99.9$ & $97.9$ & $99.4$ & $99.9$ & $100\hphantom{.0}$ & $105.3$ & $102.2$ & $101\hphantom{.0}$ & $100.1$ \\
k & $77.9$ & $95\hphantom{.0}$ & $99\hphantom{.0}$ & $100\hphantom{.0}$ & $97.4$ & $99.5$ & $100\hphantom{.0}$ & $100.1$ & $103.3$ & $100.8$ & $100.2$ & $100\hphantom{.0}$ \\
l & $138.4$ & $121.5$ & $111.7$ & $102.1$ & $111.2$ & $105\hphantom{.0}$ & $102\hphantom{.0}$ & $99.5$ & $52\hphantom{.0}$ & $59.9$ & $67.7$ & $84.8$ \\
m & $72.7$ & $89.5$ & $96.2$ & $99.8$ & $96.3$ & $98.9$ & $99.7$ & $100.1$ & $110.3$ & $104.4$ & $101.9$ & $100.2$ \\
o & $261.9$ & $125.7$ & $106.4$ & $100.3$ & $102.8$ & $100.2$ & $99.8$ & $100\hphantom{.0}$ & $69.8$ & $84.8$ & $92\hphantom{.0}$ & $98.2$ \\
p & $137.1$ & $124.6$ & $112.4$ & $101\hphantom{.0}$ & $107.1$ & $102.4$ & $100.4$ & $99.9$ & $76.6$ & $85.1$ & $91.8$ & $98.8$ \\
\end{tblr}
}}
\end{table}

 \subsection{Simulation 1}\label{sec:C1}
We consider datasets of sizes $n=2000$, and use a random grid of candidate thresholds $\calU$ at the empirical $\{0.8, 0.81, \ldots, 0.98\}$ quantiles, thus giving $20$ observations between each threshold. We compare the following threshold selection methods:
\begin{enumerate}[label={$\textsc{p}_{\arabic*}$. }]
 \item the threshold stability plot of \cite{Davison.Smith:1990}, using the smallest threshold for which point estimates for the shape are included in the profile-likelihood 95\% confidence interval for all higher thresholds;
 \item minimization of the mean squared error of the shape parameter by semiparametric bootstrap of \cite{Caers.Beirlant.Maes:1999};
 \item \cite{Gonzalo.Olmo:2004} with absolute distance, mimicking \cite{Pickands:1975};
\item \cite{Gonzalo.Olmo:2004} with the Kolmogorov--Smirnov statistic;
 \item normality tests for coefficients of \cite{Thompson:2009} ($\star$);
 \item the \cite{Suveges.Davison:2010} information matrix test with gap $K=1$ ($\star$);
 \item the score test of \cite{Northrop.Coleman:2014} comparing the piecewise generalized Pareto and generalized Pareto models ($\star$);
 \item the mean residual life plot using the automated procedure of \cite{Langousis:2016}, returning the threshold that minimizes the weighted mean squared error;
 \item the \cite{Wadsworth:2016} white noise test, returning the smallest threshold at which the white noise hypothesis cannot be rejected based on a change-point likelihood ratio test;
 \item the posterior predictive model of \cite{Northrop.Attalides.Jonathan:2017}, returning the threshold with the largest posterior weight;
 \item \cite{Northrop.Attalides.Jonathan:2017}, with Bayesian model averaging;
 \item goodness-of-fit statistics from \cite{Bader.Yan.Zhang:2018} ($\star$);
\item the \cite{SilvaLomba.FragaAlves:2020} $L$-moment skewness-kurtosis procedure;
\item \cite{Kiran.Srinivas:2021} Mahalanobis-distance minimization with $L$-moments;
 \item metric-based adjustments of \cite{Varty.Tawn:2021} with exponential quantile-quantile plots with weighted mean squared error;
 \item the likelihood ratio test to compare the extended generalized Pareto beta model of \cite{Gamet.Jalbert:2022} with the generalized Pareto ($\star$);
 \item metric-based adjustments of \cite{Murphy.Tawn.Varty:2024} with generalized Pareto quantile-quantile plots;
 \item the \cite{Collings:2025} TAILS method.
 \end{enumerate}

 All tests are performed at the nominal 5\% significance level. Procedures marked with a star ($\star$) return a sequence of $p$-values for the ordered thresholds, to which a sequential testing procedure such as ForwardStop \citep{GSell:2016} might be applied.  By design this systematically leads to the selection of lower thresholds.

 We use empirical quantiles from simulated datasets to ensure the number of exceedances stays the same, rather than fixing the thresholds to the true quantiles of the distribution. The latter would facilitate comparisons, but could not be applied in practice.

We also devise an oracle that minimizes the absolute bias among all candidate thresholds; only maximum likelihood estimates evaluated at empirical quantiles in the range \{0.8, 0.81, \ldots, 0.98\} are considered; the loss is estimated using a Monte Carlo average with 1000 samples from each scenario, excluding missing values due to failed convergence, or when methods do not return a threshold.

We also checked the effect of varying the sample size, taking $n \in \{1000, 2000, 3000, 4000\}$ and retaining the same quantile level as candidate thresholds, adding the 0.99 quantile for the two largest values of $n$. We also considered scenarios with serial dependence and rounding. For the first, time series were drawn from a Gaussian autoregressive process of order 1 with correlation coefficient 0.5, and transformed to have the marginal distribution of interest using the probability integral transform. For the rounding, we multiplied each series by a factor equal to one hundred divided by the 0.9 quantile of the distribution, $c =100/F^{-1}(0.9)$, rounded observations up to the nearest integer to avoid creating zeros, then back-transformed them , viz. $x \mapsto c\lceil x/c\rceil$. This leads to more rounding at lower quantile levels than higher ones.

%

\subsection*{Summary of results for Simulation 1}

\begin{table}
 \centering
\caption{Median ratio of absolute difference to the true quantile versus oracle $|q_{\text{method}}-q_0|/|q_{\text{oracle}} - q_0|$ for different estimation methods (rows) and different distributions (columns) for the estimation of 0.999 quantile for Simulation 1. Smaller numbers are preferable}
\begin{tabular}{lrrrrrrrrrrrrrrrr}
\toprule
 & a & b & c & d & e & f & g & h & i & j & k & l & m & n & o & p\\
\midrule
$\textsc{p}_{1}$ & $1.6$ & $2.3$ & $2.4$ & $1.5$ & $1.4$ & $1.7$ & $2.0$ & $1.5$ & $1.8$ & $2.8$ & $3.1$ & $3.6$ & $2.4$ & $4.0$ & $2.7$ & $3.4$\\
$\textsc{p}_{2}$ & $1.5$ & $2.4$ & $2.3$ & $1.5$ & $1.3$ & $1.7$ & $2.0$ & $1.4$ & $1.7$ & $2.6$ & $3.1$ & $3.9$ & $2.3$ & $4.0$ & $2.6$ & $3.2$\\
$\textsc{p}_{3}$ & $1.7$ & $2.7$ & $2.6$ & $1.6$ & $1.4$ & $1.9$ & $2.1$ & $1.5$ & $1.9$ & $3.0$ & $3.5$ & $4.7$ & $2.6$ & $3.8$ & $2.5$ & $3.4$\\
$\textsc{p}_{4}$ & $1.9$ & $2.8$ & $2.7$ & $1.7$ & $1.4$ & $2.0$ & $2.4$ & $1.6$ & $2.2$ & $3.6$ & $4.8$ & $5.0$ & $3.2$ & $4.0$ & $2.5$ & $3.5$\\
$\textsc{p}_{5}$ & $1.6$ & $2.6$ & $2.6$ & $1.6$ & $1.4$ & $1.9$ & $2.2$ & $1.5$ & $1.9$ & $3.1$ & $3.8$ & $4.7$ & $2.6$ & $3.8$ & $2.6$ & $3.4$\\
$\textsc{p}_{6}$ & $1.7$ & $2.6$ & $2.4$ & $1.6$ & $1.4$ & $1.8$ & $2.2$ & $1.5$ & $1.9$ & $3.1$ & $3.7$ & $4.1$ & $2.6$ & $4.0$ & $2.4$ & $3.2$\\
$\textsc{p}_{7}$ & $1.9$ & $2.8$ & $2.5$ & $1.8$ & $1.4$ & $1.9$ & $2.5$ & $1.6$ & $2.1$ & $3.6$ & $4.6$ & $5.4$ & $3.4$ & $4.2$ & $2.4$ & $3.5$\\
$\textsc{p}_{8}$ & $1.5$ & $2.2$ & $2.2$ & $1.5$ & $1.3$ & $1.6$ & $2.0$ & $1.4$ & $1.8$ & $2.6$ & $2.9$ & $3.6$ & $2.3$ & $3.8$ & $2.7$ & $3.3$\\
$\textsc{p}_{9}$ & $1.7$ & $2.4$ & $2.5$ & $1.6$ & $1.4$ & $1.7$ & $2.1$ & $1.6$ & $1.9$ & $2.7$ & $3.0$ & $3.8$ & $2.4$ & $3.8$ & $2.4$ & $3.4$\\
$\textsc{p}_{10}$ & $1.5$ & $2.3$ & $1.9$ & $1.5$ & $1.3$ & $1.6$ & $1.9$ & $1.3$ & $1.7$ & $2.7$ & $3.1$ & $3.5$ & $2.5$ & $4.1$ & $1.7$ & $2.6$\\
$\textsc{p}_{11}$ & $1.8$ & $2.7$ & $2.5$ & $1.7$ & $1.4$ & $2.0$ & $2.5$ & $1.6$ & $2.2$ & $3.4$ & $4.4$ & $4.6$ & $3.0$ & $4.1$ & $2.6$ & $3.4$\\
$\textsc{p}_{12}$ & $1.8$ & $2.8$ & $2.6$ & $1.7$ & $1.5$ & $2.0$ & $2.3$ & $1.6$ & $2.1$ & $3.3$ & $4.1$ & $4.5$ & $3.0$ & $3.9$ & $2.6$ & $3.4$\\
$\textsc{p}_{13}$ & $1.7$ & $2.5$ & $2.4$ & $1.6$ & $1.3$ & $1.7$ & $2.1$ & $1.5$ & $2.0$ & $2.9$ & $3.5$ & $4.4$ & $2.7$ & $3.8$ & $2.6$ & $3.4$\\
$\textsc{p}_{14}$ & $1.9$ & $3.2$ & $2.8$ & $1.8$ & $1.4$ & $2.0$ & $2.5$ & $1.6$ & $2.2$ & $4.0$ & $4.0$ & $4.0$ & $3.5$ & $4.7$ & $2.4$ & $3.8$\\
$\textsc{p}_{15}$ & $1.7$ & $2.8$ & $2.4$ & $1.6$ & $1.4$ & $2.0$ & $2.4$ & $1.5$ & $2.1$ & $3.3$ & $3.4$ & $4.1$ & $3.0$ & $3.8$ & $2.5$ & $3.4$\\
$\textsc{p}_{16}$ & $1.6$ & $2.5$ & $2.3$ & $1.6$ & $1.4$ & $1.7$ & $2.0$ & $1.5$ & $1.9$ & $2.8$ & $3.2$ & $4.1$ & $2.5$ & $3.8$ & $2.6$ & $3.2$\\
$\textsc{p}_{17}$ & $1.7$ & $2.5$ & $2.5$ & $1.6$ & $1.4$ & $1.8$ & $2.1$ & $1.5$ & $2.1$ & $3.3$ & $3.9$ & $4.6$ & $2.8$ & $3.8$ & $2.4$ & $3.6$\\
$\textsc{p}_{18}$ & $1.6$ & $2.4$ & $2.3$ & $1.6$ & $1.4$ & $1.7$ & $2.0$ & $1.5$ & $1.8$ & $2.7$ & $3.1$ & $3.9$ & $2.3$ & $3.5$ & $2.7$ & $3.1$\\
$\textsc{bma}$ & $1.6$ & $2.4$ & $2.4$ & $1.6$ & $1.4$ & $1.8$ & $2.0$ & $1.5$ & $1.8$ & $2.7$ & $3.1$ & $3.9$ & $2.3$ & $3.6$ & $2.7$ & $3.4$\\
\bottomrule
\end{tabular}
\label{tab:vsoracle}
\end{table}

\begin{figure}[t!]
\centering
\includegraphics[width = \textwidth]{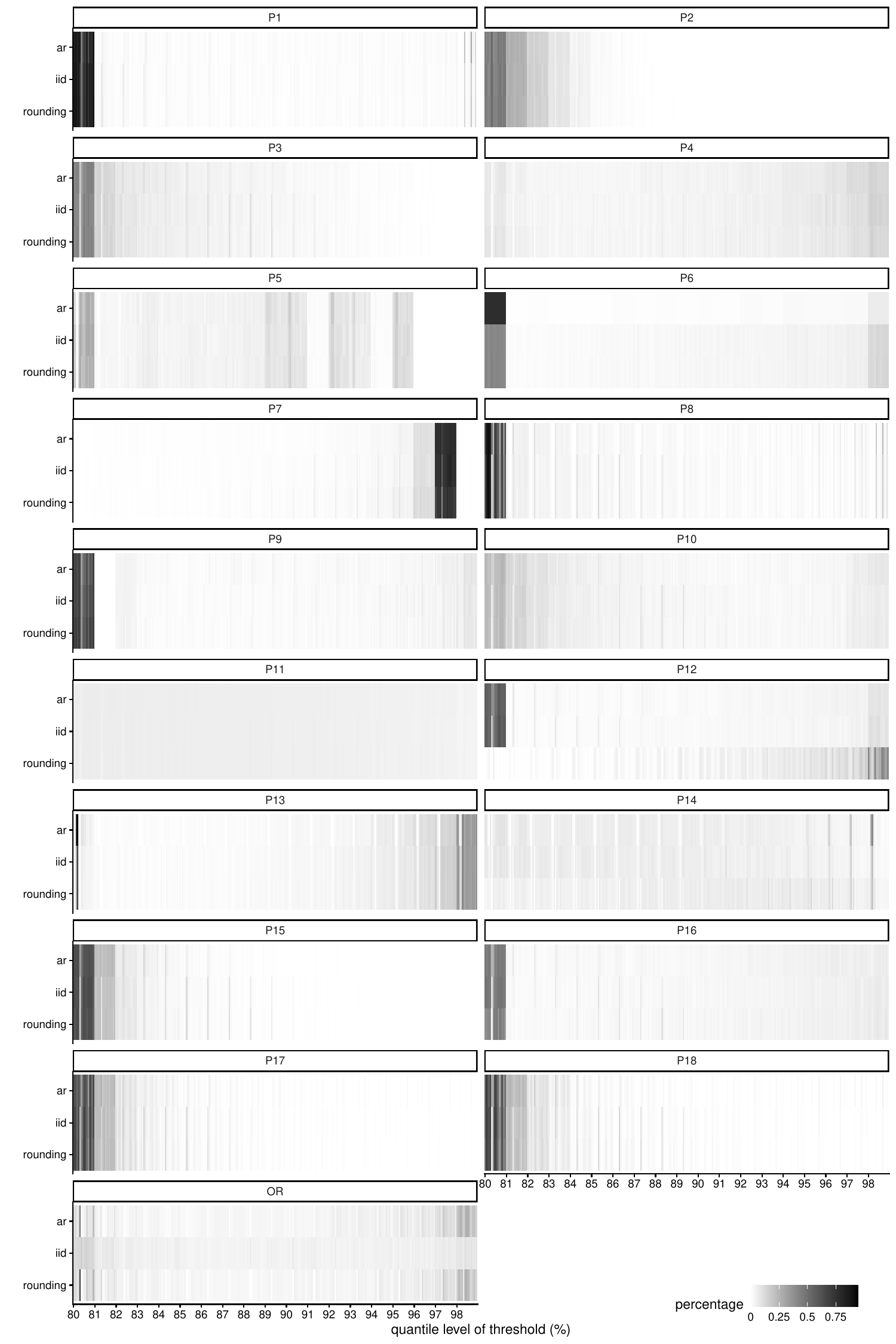}
 \caption{Heatmap of the percentage of threshold selection  per method for each of the different quantile levels ($x$-axis, from left to right), across scenarios ($y$-axis) and distributions a--p (panels).}
 \label{fig:heatmap_thselect_simu1}
\end{figure}
\begin{figure}[th!]
\centering
\includegraphics[width=\textwidth]{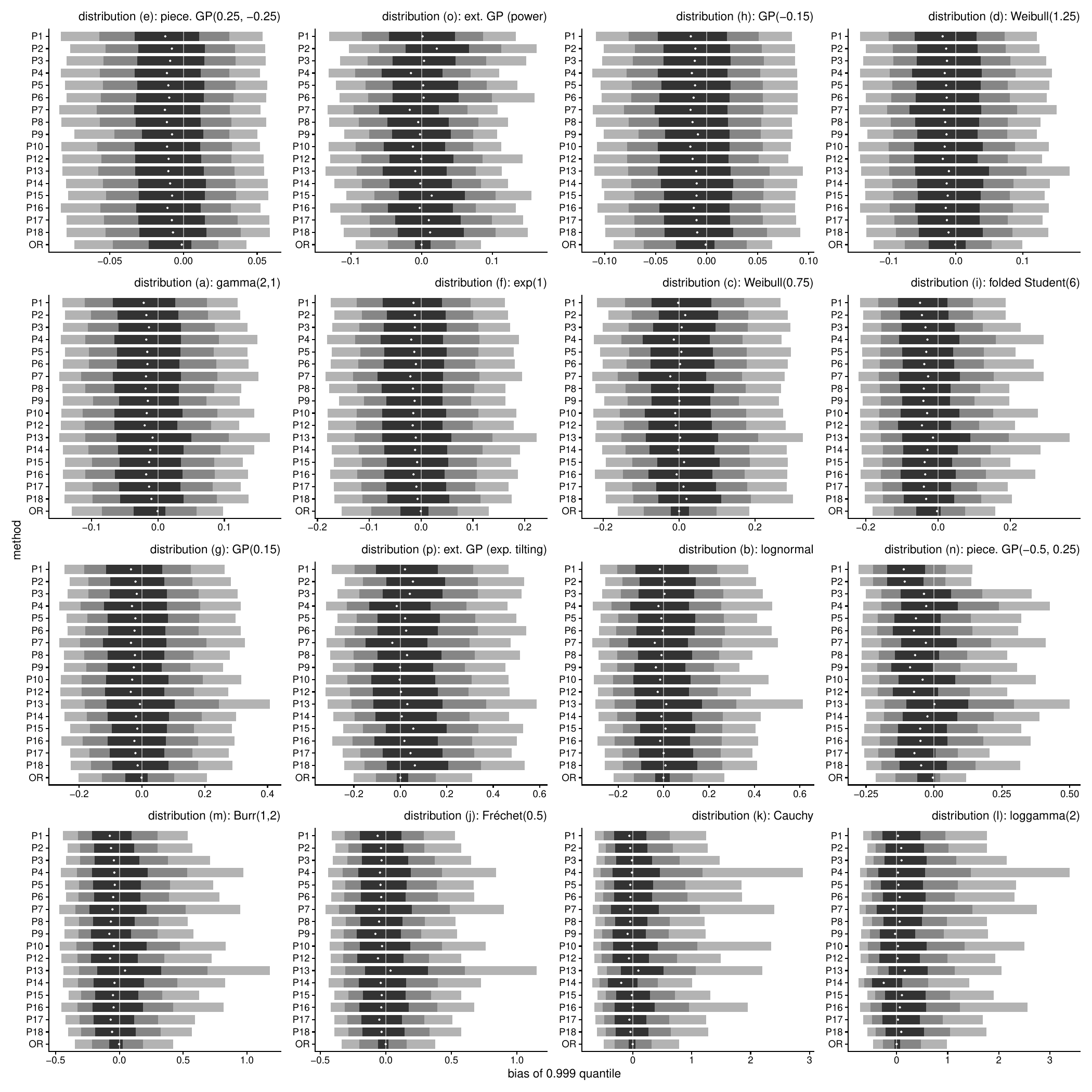}
\caption[Simulation study 1 results]{Results of Simulation 1: median (white circles), 50\% (dark grey), 80\% (grey) and 95\% (light grey) bands for the relative bias of the 0.999 quantile, based on maximum likelihood estimates using candidate thresholds at the $0.8, 0.81, \ldots, 0.98$ sample quantiles taken from $n=2000$ independent and identically distributed observations. The distributions are ordered in increasing level of the penultimate shape parameter (vertical black line) at level 0.999.}
\label{figsim2a}
\end{figure}

Give the number of distributions and methods considered, it is difficult to summarize results. We consider first in \Cref{fig:heatmap_thselect_simu1} the proportion of time a quantile level is returned as threshold (as a function of rounding and truncation), as this is indicative of the sensibility of the method to the underlying data generating process and reveals some methods for which universally, the lowest candidate threshold seems to be favored.

Summary statistics of the sampling distribution of the estimated bias of the 0.999 quantile (relative to the true 0.999 quantile), $(\widehat{q}_{0.999}-q_0)/q_0$ are shown in \Cref{figsim2a}. A lot of the performance has to do with the maximum likelihood estimator, and many distributions have similar performances because the threshold selection methods return the same (low) threshold levels; see \Cref{fig:heatmap_thselect_simu1}. 
The median ration of these absolute biases, relative to the oracle, are reported in \Cref{tab:vsoracle}.

Increasing the sample size had little effect on choice: most methods display changes of the order 1--2\%, so are globally insensitive to sample size. The methods most affected were $\textsc{p}_{4}$, $\textsc{p}_{6}$ and $\textsc{p}_{13}$, which saw an increase in the proportion of higher thresholds retained. Method $\textsc{p}_{8}$ displayed the reverse tendency, suggesting that the weighting scheme is wholly inadequate. Methods $\textsc{p}_{5}$ and $\textsc{p}_{7}$, on the other hand, are sensitive to the choice of alternatives and the number of thresholds.

In all cases, ignoring the serial dependence leads to large positive relative biases of the maximum likelihood estimators of quantiles or return levels. Truncation, however, has little to no impact.
The information matrix test of $\textsc{p}_{6}$, designed for joint estimation of the threshold and extremal index, shows a noticeable increase in the proportion of lowest thresholds chosen, from roughly 40\% in the iid case to 77\% for the autoregressive model. Method $\textsc{p}_{8}$ saw a more modest increase of 5\% in the frequency at which the 0.8 quantile was chosen.

Method $\textsc{p}_{12}$ is extremely sensitive to rounding, with average differences of around 57\% in the proportion of times the 0.8 quantile was chosen as threshold. The reason for this is that the null distribution in the software implementation is obtained by calculating the goodness-of-fit measure based on simulated samples, which are not rounded. Introducing rounding in the parametric Monte Carlo approximation to $p$-values would solve this. Methods $\textsc{p}_{9}$, $\textsc{p}_{14}$, $\textsc{p}_{15}$ and $\textsc{p}_{17}$ are also affected, to a lesser degree, especially for the Weibull distribution (d).
\begin{itemize}
\item The modal quantile level for many procedures is 0.8, the lowest candidate. The procedures of \cite{Thompson:2009} ($\textsc{p}_5$), \cite{Northrop.Coleman:2014} ($\textsc{p}_7$) and \cite{SilvaLomba.FragaAlves:2020} ($\textsc{p}_{13}$) select higher thresholds.
\item Application of ForwardStop (not shown) leads to lower thresholds and is thus is not recommended, as most methods generally give low thresholds.
\item The methods of \cite{Thompson:2009} ($\textsc{p}_{5}$), \citet{Suveges.Davison:2010} ($\textsc{p}_{6}$), \citet{Northrop.Attalides.Jonathan:2017} ($\textsc{p}_{11}$), \citet{SilvaLomba.FragaAlves:2020} ($\textsc{p}_{13}$) and \citet{Kiran.Srinivas:2021} ($\textsc{p}_{14}$) lead to much less agreement and greater variability of selected quantile levels.
For $\textsc{p}_{11}$, this tallies with Table~2 of \citet{Murphy.Tawn.Varty:2024}.

\item \citet{Wadsworth:2016} ($\textsc{p}_{9}$) sequential testing fails 33\% of the time, even after reducing the number of thresholds considered, jittering data, and with sample sizes of 1000 observations. It performs worst with heavy-tailed distributions, but is generally competitive when it works.
 \item The procedures of \cite{Caers.Beirlant.Maes:1999} ($\textsc{p}_{2}$), \citet{Wadsworth:2016} ($\textsc{p}_{9}$) and \cite{Murphy.Tawn.Varty:2024} ($\textsc{p}_{17}$) perform well relative to the oracle.
 \item By construction, the \citet{Collings:2025} TAILS method ($\textsc{p}_{18}$) selects higher thresholds than \citet{Murphy.Tawn.Varty:2024} ($\textsc{p}_{17}$) and thus performs better than the latter for heavy-tailed distributions. Both methods outperform \citet{Varty.Tawn:2021} ($\textsc{p}_{16}$).
 \item The Bayesian model averaging method of \cite{Northrop.Attalides.Jonathan:2017} is competitive in all scenarios. In our simulation, the posterior weights of each quantile levels tended to be more or less uniformly distributed over the set of candidates, with less weight for the largest quantiles.
\end{itemize}

\subsection{Simulation 2}
Many of the semiparametric methods are not directly comparable with their likelihood-based counterparts because they choose an order statistic among all possible choices, rather than a threshold from a candidate set. Most such methods (except $\textsc{s}_{3}$ and $\textsc{s}_{14}$) are based on Hill's estimator, and designed for positive shape parameters, a major practical limitation. Moreover, most methods are designed to provide good point estimators of the shape, which are seldom of interest \emph{per se}; there is no guarantee that these properties carry over to risk measures.

We compared the following procedures, most of which are implemented in the \textbf{R} packages \texttt{tea} \citep{tea} and \texttt{mev} \citep{mev}:
\begin{enumerate}[label={$\textsc{s}_{\arabic*}$.}]
\item minimization of the asymptotic mean squared error of the Hill estimator \citep{Hall.Welsh:1985};
\item smoothing and bootstrap estimation of the mean squared error \citep{Hall:1990};
\item the exponential generalized quantile threshold of \cite{Beirlant.Vynckier.Teugels:1996a};
\item the bias-reduction method of \cite{Drees.Kaufmann:1998};
\item minimization of the asymptotic mean squared error of the Hill estimator, estimated using a nonparametric double bootstrap \citep{Danielsson:2001};
\item the bootstrap diagnostic test for exponentiality of log-spacings \citep{Guillou.Hall:2001};
\item the non-robust prediction error $C$-criterion (non-robust version) of \cite{Dupuis.Victoria-Feser:2003};
\item minimization of \citep[][p.~137]{Reiss.Thomas:2007}
\begin{align*}
\frac{1}{n_u} \sum_{i=n-n_u}^{n} i^\beta \left|H_{n,i} - \mathrm{med}\{H_{n,n}, \ldots, H_{n, i+1}\}\right|^p, \qquad 0 \leq \beta < \frac{1}{2},
\end{align*}
for $p=1$, with $\beta=0$ based on recommendations from \cite{Neves:FragaAlves:2004} for Hill's estimator with heavy-tailed data;
\item \cite{Reiss.Thomas:2007}, as above but with $p=2$;
\item Jackson kernel-based threshold selection  \citep{Goegebeur.Beirlant.deWet:2008};
\item minimum distance threshold selection \citep{Clauset.Shalizi.Newman:2009};
\item minimization of the asymptotic mean squared error of the Hill estimator, estimated using a double nonparametric bootstrap scheme \citep{Gomes.Figueiredo.Neves:2012};
\item a heuristic algorithm based on sample path stability \citep{Gomes:2013};
\item the random block maxima estimator of \cite{Wager:2014} with empirical risk minimization;
\item a variant of \cite{Hall:1990} that estimates second-order parameters \citep{Caeiro.Gomes:2014};
\item minimization of the asymptotic mean squared error of the Hill estimator \citep[][\S~2]{Caeiro.Gomes:2016};
\item the eyeballing technique of \cite{Danielsson:2019} based on moving windows for the Hill plot;
\item minimization of the mean absolute deviation between the largest observations of the dataset and the theoretical generalized Pareto tail \citep{Danielsson:2019}, estimated using Hill's estimator;
\item the procedure of \cite{Danielsson:2019}, but with the Kolmogorov--Smirnov distance;
\item minimization of the asymptotic mean squared error based on the relationship between Hill and trimmed Hill estimators of \cite{Bladt:2020};
\item smooth estimation of the asymptotic mean squared error of the generalized jackknife estimator (SAMSEE) of \cite{Schneider:2021}.
\end{enumerate}
In each case we return the selected threshold, the number of exceedances, the shape parameter estimate and \citeauthor{Weissman:1978}'s quantile estimate at level $0.999$. All methods but $\textsc{s}_{3}$ and $\textsc{s}_{14}$ use the Hill estimator. The exponential regression model shape estimator can be negative and, although the estimator described in Remark~2 of \cite{Beirlant.Dierckx.Guillou:2005} could be used, results would not be easily comparable. As such, quantile are omitted for this method.

\begin{figure}[th!]
\centering
\includegraphics[width=\textwidth]{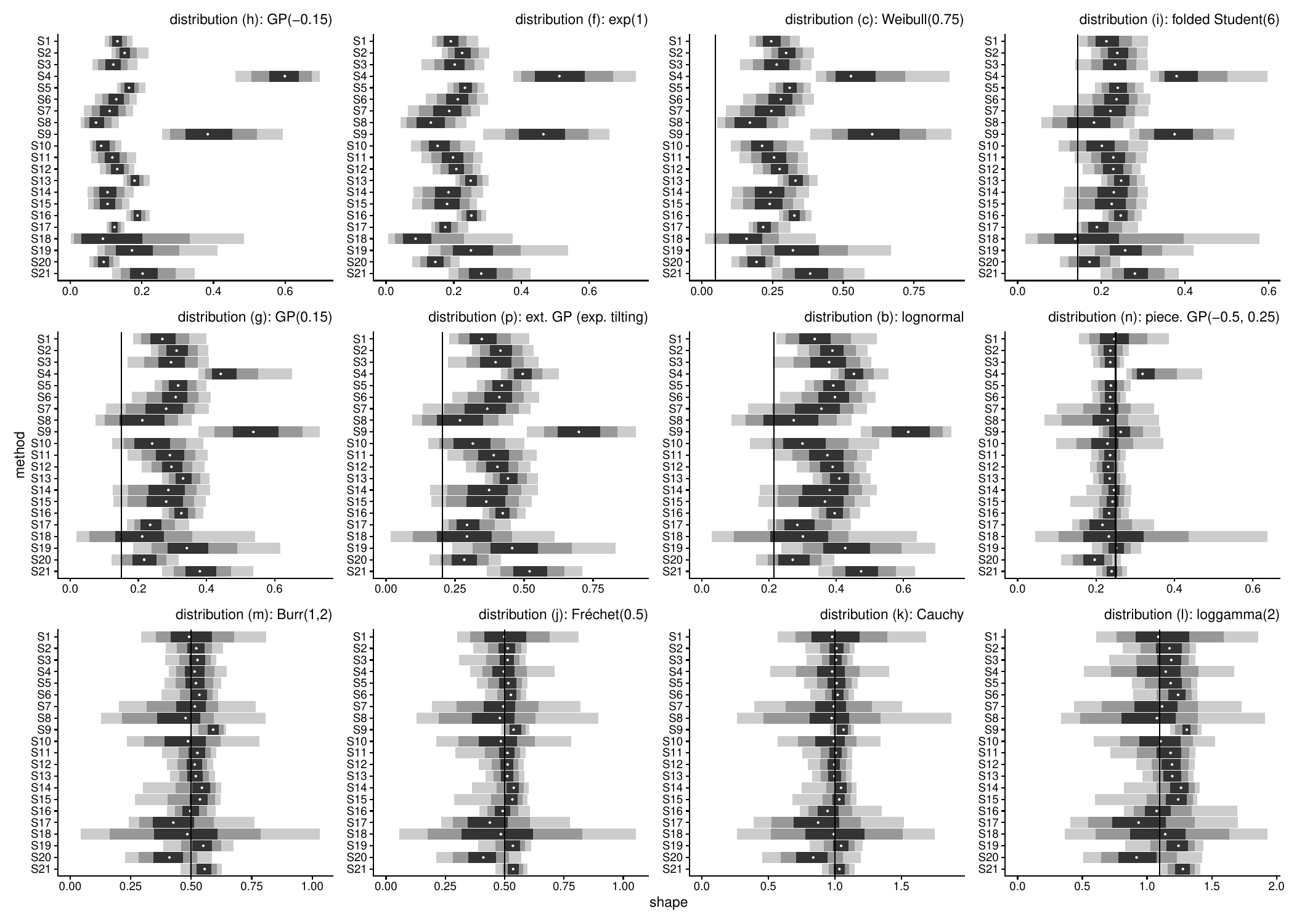}
\caption[Simulation 2 results]{Results of Simulation 2: density plots with median, 66\% and 95\% bands for the shape parameter estimates (top), as a function of the distribution, method for samples of $n=2000$ independent and identically distributed observations. The distributions are ordered by increasing level of the penultimate shape parameter (vertical black line).}
\label{figsim2a}
\end{figure}
\begin{figure}[th!]
\centering
\includegraphics[width=\textwidth]{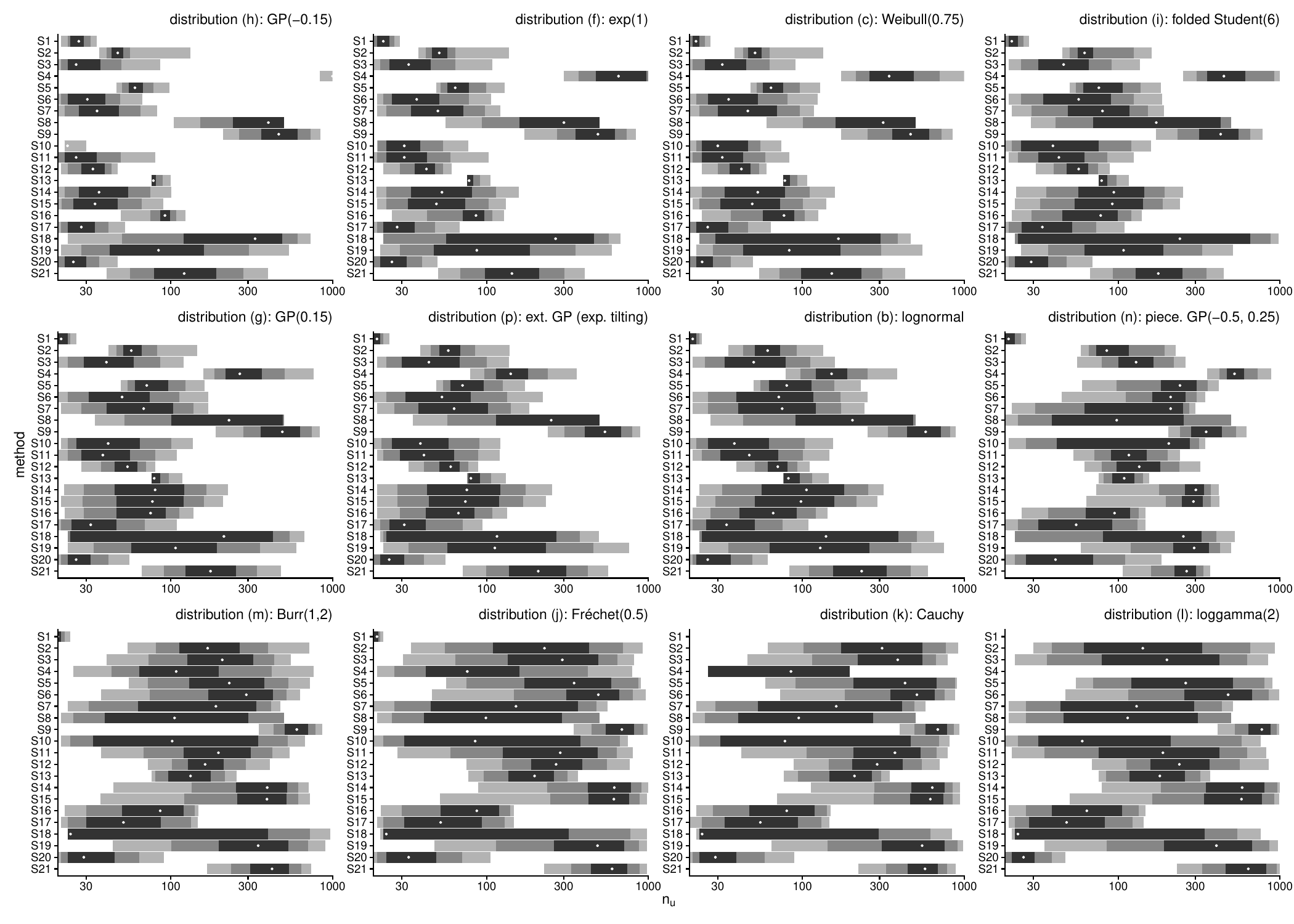}
\caption[Simulation 2 results]{Results for Simulation 2: median (white point) 50\% (dark grey), 80\% (grey) and 95\% (light grey) confidence bands for the rank of the threshold (with the number of exceedances on the $x$-axis shown on the log-scale), chosen as the $(n-n_u)$th order statistic with the same settings of \Cref{figsim2a}.}
\label{figsim2b}
\end{figure}
%

\subsection*{Robustness of semiparametric methods}

We performed a simulation similar to that of Simulation~1  for rounding and serial correlation, and for determining the effect of sample size on estimates. We focus on generalized Pareto distributions, and those for which the penultimate shape parameter is positive.

\subsubsection*{Impacts of rounding and truncation}

\begin{figure}[th!]
\centering
\includegraphics[width=\textwidth]{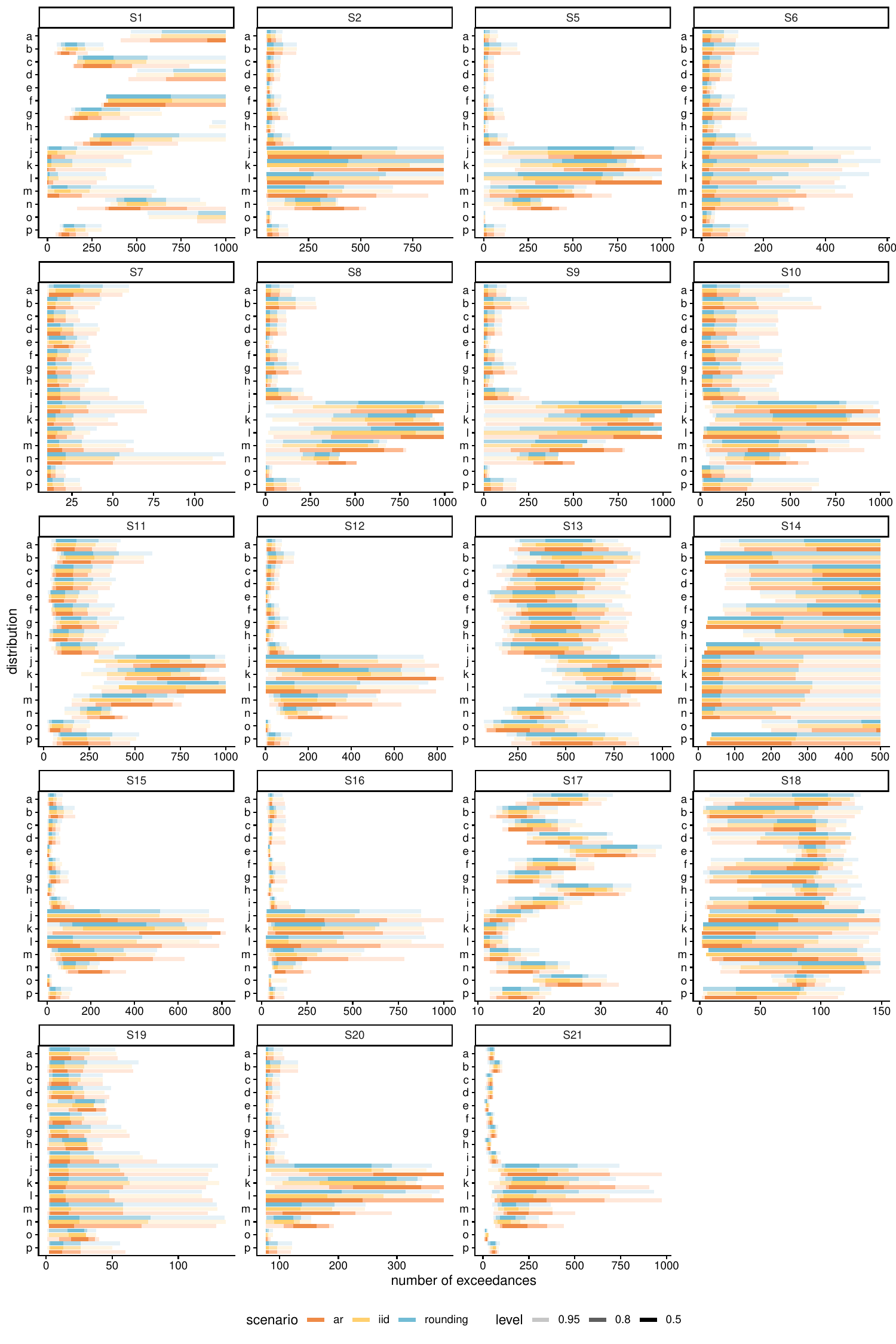}
\caption[Simulation 2 results]{Results for Simulation 2: 50\%, 80\% and 95\% (shade) highest density continuous intervals for the number of exceedances for different scenarios (independent and identically distributed, serially correlated and rounded).}
\label{figsim2h}
\end{figure}

\begin{figure}[th!]
\centering
\includegraphics[width=\textwidth]{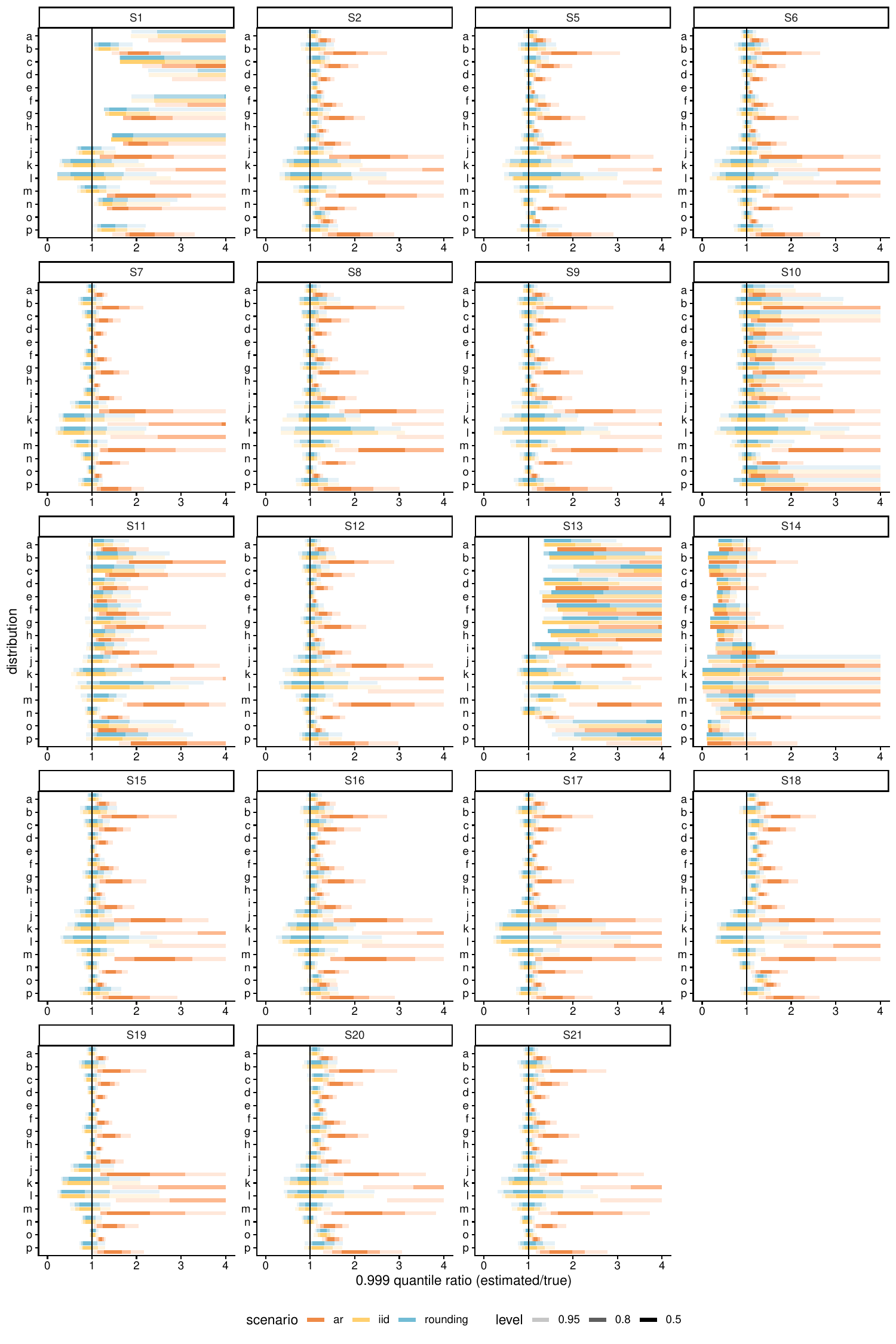}
\caption[Simulation 2 results]{Results for Simulation 2: 50\%, 80\% and 95\% highest density continuous intervals for the ratio of the 0.999 quantile estimate, relative to the true theoretical counterpart.}
\label{figsim2i}
\end{figure}

The effect of rounding was not particularly noticeable on the quantile level of the thresholds (\Cref{figsim2h}). On the contrary, serial correlation generally leads to the selection of lower quantiles for heavier-tailed distributions. Failing to account for dependence leads to positively biased shape estimators and thus to fairly systematic overestimation of high quantiles for heavy-tailed distributions (\Cref{figsim2i}). Methods $\textsc{s}_{3}$ and $\textsc{s}_{14}$, which use different shape estimators, appear more robust to autocorrelation.

\subsubsection*{Sample size}
We consider two different schemes for assessing the impact of sample size and to see whether, in larger samples, increasingly high quantile levels are selected as thresholds.

\begin{figure}[th!]
\centering
\includegraphics[width=\textwidth]{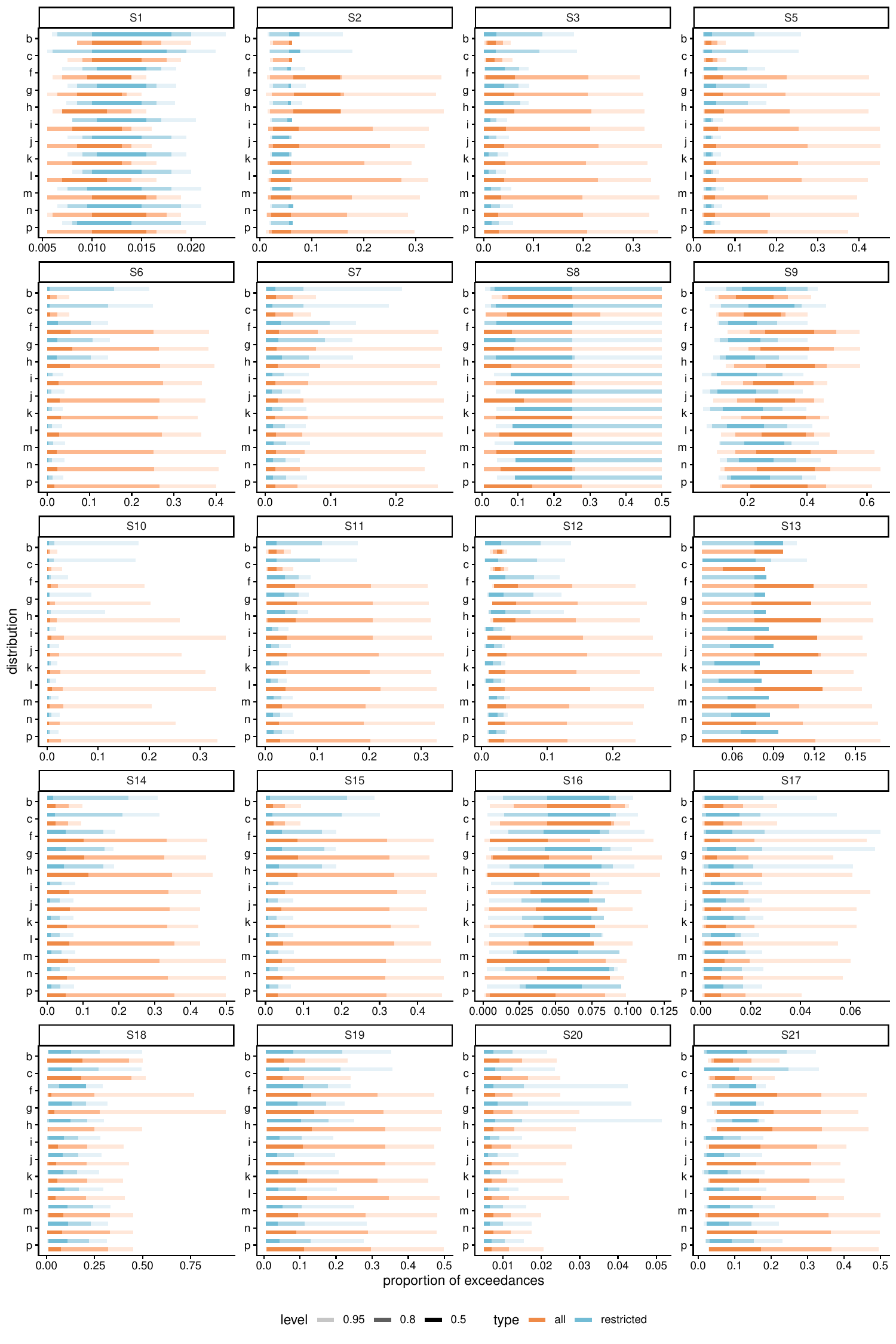}
\caption[Simulation 2 results]{Results for Simulation 2: 50\%, 80\% and 95\% highest density continuous intervals for the proportion of exceedances relative to the sample size for $n=2000$, when restricting the largest 50\% or not (color).}
\label{figsim2e}
\end{figure}

\begin{figure}[th!]
\centering
\includegraphics[width=\textwidth]{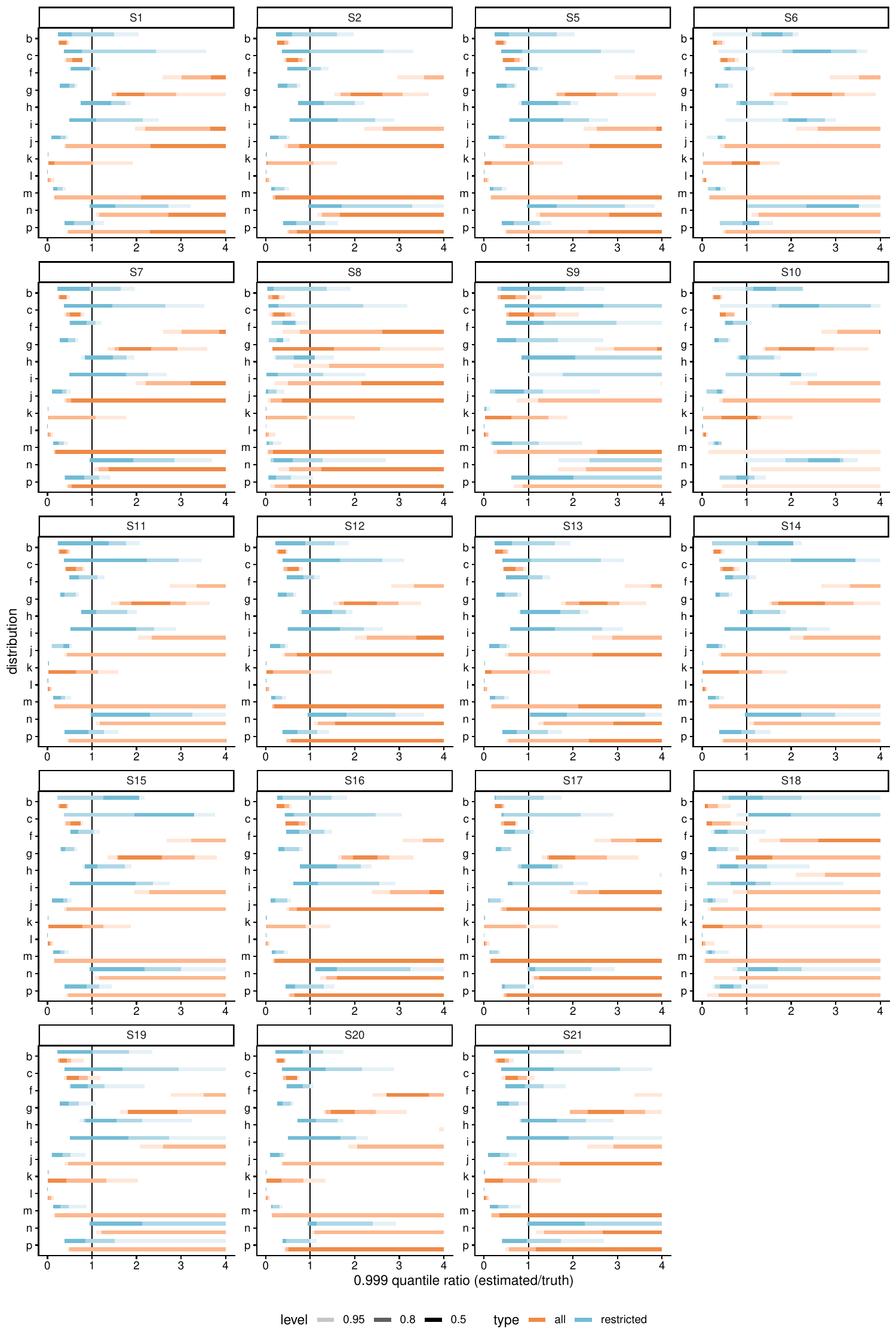}
\caption[Simulation 2 results]{Results for Simulation 2: 50\%, 80\% and 95\% highest density continuous intervals for the ratio of estimated 0.999 quantile to the true theoretical counterpart, when restricting the largest 50\% or not (color). The ratio is capped at 4 in the plot.}
\label{figsim2f}
\end{figure}

\begin{figure}[th!]
\centering
\includegraphics[width=\textwidth]{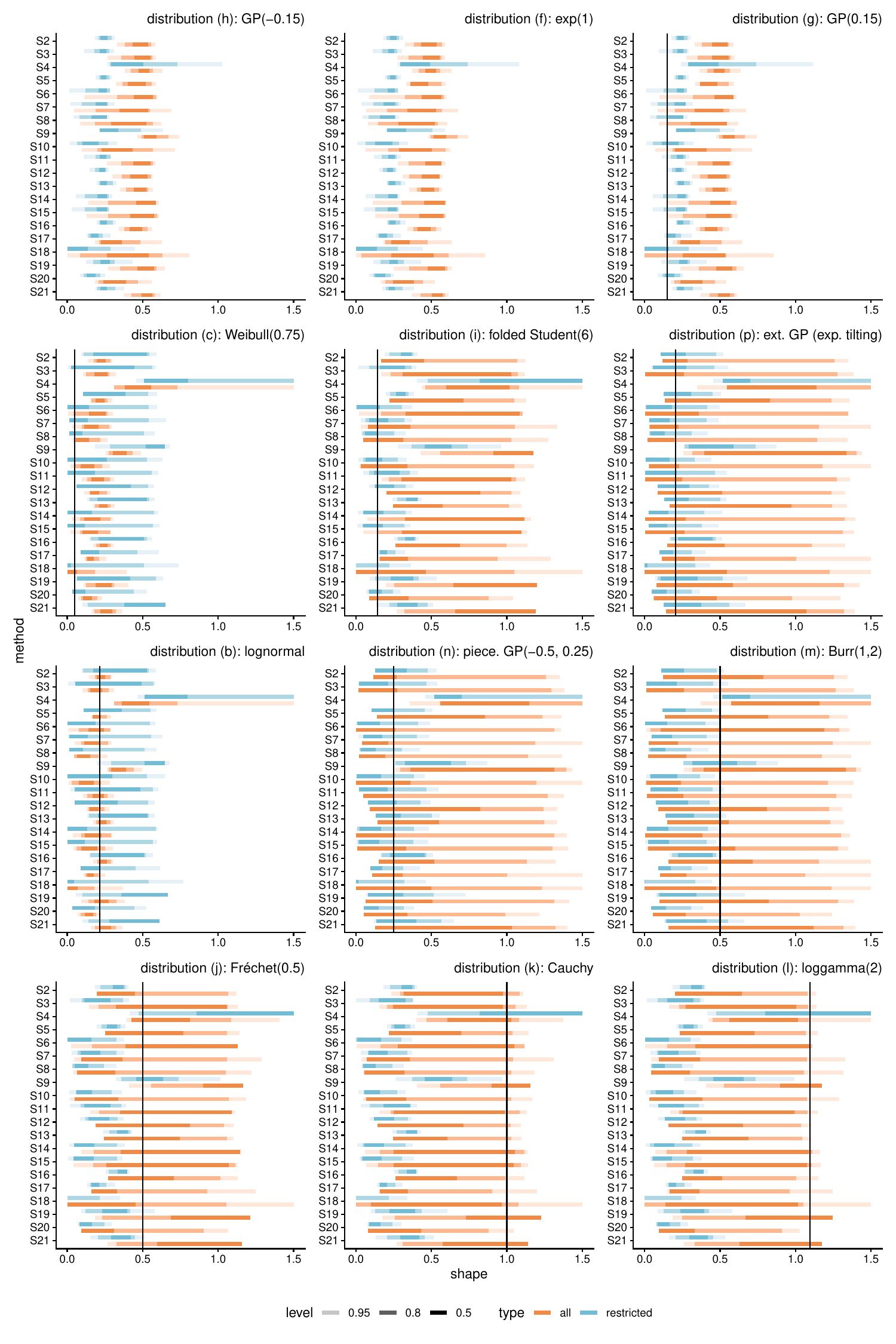}
\caption[Simulation 2 results]{Results for Simulation 2: 50\%, 80\% and 95\% highest density continuous intervals for the shape parameter estimates for generalized Pareto models (top row) and models with penultimate approximations and heavy tails. The vertical bar in each panel gives the value of the penultimate shape parameter at the 0.999 quantile.}
\label{figsim2g}
\end{figure}

In the first scheme, we consider samples of size $n=2000$, and consider the effect of restricting selection to $n=1000$ top observations (i.e., above the sample median) or allowing all $n=2000$ as candidate thresholds. \Cref{figsim2e} shows the distribution of the proportion of exceedances, $n_u/n$, per method and over all distribution scenarios. It shows quantiles of the density, truncated to the range $[0.1, 0.9]$, for the unconstrained case (all) and the restricted case. Many methods are very sensitive to the non-extreme observations, and for most the range of selected values for $n_u$ becomes both more variable and wider when non-extreme observations are selected as thresholds. Methods $\textsc{s}_{2}$, $\textsc{s}_{3}$, $\textsc{s}_{5}$--$\textsc{s}_{7}$, $\textsc{s}_{11}$--$\textsc{s}_{15}$ and $\textsc{s}_{21}$ all see a much wider range for the distribution of $n_u/n$ when unrestricted, though the opposite can occur with light-tailed distributions.
Generally, selecting lower thresholds (larger $n_u$) in these heavy-tailed scenarios leads to a large increase in the shape parameter estimates (\Cref{figsim2g}) which also become much more dispersed, biasing the estimated quantiles (\Cref{figsim2f}) and increasing the variability of the sampling distribution of the parameter estimators.

\begin{figure}[th!]
\centering
\includegraphics[width=\textwidth]{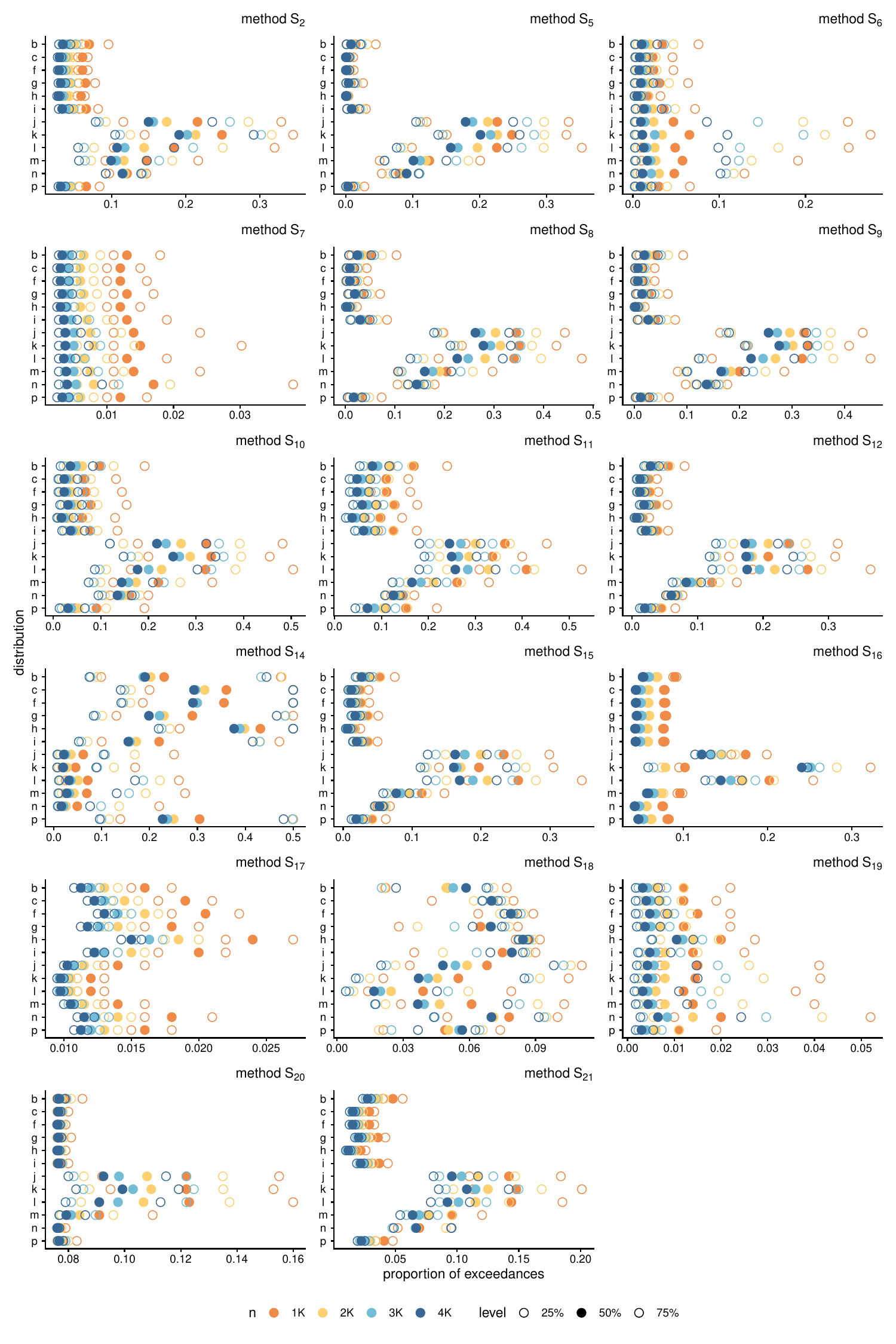}
\caption[Simulation 2 results]{Results for Simulation 2: quartiles and median (full circle) of the percentage of exceedances relative to the total sample size $n \in \{1000, 2000, 3000, 4000\}$ ($x$-axis) for each distribution ($y$-axis) and for selected threshold selection methods (panel), as a function of the sample size (color).}
\label{figsim2d}
\end{figure}

In the second scheme, we consider samples of sizes $n=1000, 2000, 3000, 4000$. For most methods and distributions, and as one would hope, the quantile level of the threshold increased with $n$. Methods $\textsc{s}_{17}$ and $\textsc{s}_{19}$ select very high quantiles (higher than the 0.98 quantile), and likewise for $\textsc{s}_{21}$ for non-heavy tailed distributions. Some methods such as $\textsc{s}_{20}$ \citep{Bladt:2020} and the MAD minimization of \cite{Danielsson:2019} ($\textsc{s}_{18}$) had near constant proportions of exceedances, while this was the case only for some distributions for $\textsc{s}_{5}$. \Cref{figsim2d} shows how this proportion generally decreases, even if $n_u$ increases.

\subsection*{Summary of findings for Simulation 2}
\Cref{figsim2a,figsim2b} suggest the following:
\begin{itemize}
\item The Hill estimator struggles with distributions for which $\xi$ is low, but excels for very heavy-tailed distributions. The exponential estimator and the random block maxima are more variable for the latter case.
\item Estimates of the 0.999 quantile are more or less adequate even when the Hill estimator is inapplicable, at least if the penultimate behaviour of the distribution is positive.
\item The \cite{Beirlant.Vynckier.Teugels:1996b} procedure ($\textsc{s}_3$) is extremely variable, both for the selection of $n_u$ and for the estimation of $\xi$. It often fails to fit.
\item The methods of \cite{Hall.Welsh:1985} ($\textsc{s}_{1}$) and \cite{Caeiro.Gomes:2014} ($\textsc{s}_{13}$) behave erratically with small shape parameters, giving shape parameter estimates that show strong upward bias. They retain more than 15\% of the data for inference, and thus have a tendency to systematically overestimate high quantiles. For method $\textsc{s}_{13}$, the median of the ratio of 0.999 to the true quantile is about three.
\item The proposal of \cite{Schneider:2021} ($\textsc{s}_{21}$) works well in the heavy-tailed case;
\item The AMSE minimisation procedures of \cite{Caeiro.Gomes:2016} and \cite{Gomes.Figueiredo.Neves:2012} display low relative root mean squared error for quantile estimation, consistently across all scenarios, 
but can fail catastrophically for particular datasets.
\item The method with the lowest relative bias for quantiles is \cite{Guillou.Hall:2001}, and the lowest mean square errors are obtained using the \cite{Danielsson:2019} minimization of the Kolmogorov--Smirnov distance (for $\xi < 0.25$) and \cite{Bladt:2020} ($\xi \geq 0.5$);
\item The \cite{Drees.Kaufmann:1998} procedure fails to return valid values for $n_u$ in around 80\% of cases. When it works, it leads to small values of $n_u$, and to shape estimates that are below average; the width of the confidence intervals in \Cref{figsim2a} suggests excess variability. We have not fixed the tuning parameter of the stopping criterion to unity throughout.
\item The \cite{Dupuis.Victoria-Feser:2003} estimator leads to very small $n_u$, and thus variable and negatively-biased shape estimates, but performs best for quantile estimation for low shape parameter.
\item The sampling distributions of the shape parameters for the methods of \cite{Reiss.Thomas:2007} ($\textsc{s}_{8}$ and $\textsc{s}_{9}$) are left-skewed.
\item The procedure of \cite{Wager:2014} ($\textsc{s}_{14}$) provides only an estimate of the shape parameter, and performs very poorly when coupled with Weissman's estimator, with systematic underestimation of return levels; more work is needed to derive a suitable estimator of the quantile.
\end{itemize}
The more computationally-intensive methods use the bootstrap. The double bootstrap takes over two minutes to run with 1000 exceedances.

\section{Applications to data}\label{sec:dataapp}

We considered threshold selection methods for four datasets commonly used in the literature for comparison purposes. These are
\begin{enumerate}
 \item the River Nidd levels in Yorkshire 
[$n=154$, \texttt{nidd.thresh}, \texttt{evir} package],
\item the Danish fire loss [$n=2167$, \texttt{danish}, \texttt{evir} package],
\item 
the Fort Collins precipitation series [$n=4147$, \texttt{FCwx}, \texttt{extRemes} package] 
\item the North Sea wave height hindcast data [$n=628$, \texttt{ns}, \texttt{threshr} package].
\end{enumerate}
For the Fort Collins data, we consider only rainfall between April and November to avoid glaring nonstationarity.

The results are shown in \Cref{tab:real-data}. The first three datasets exhibit heavy tails, while the wave height dataset clearly has a bounded upper tail.
Many algorithms for semiparametric methods select very low thresholds for which there are a larger number of exceedances; only providing the largest 1000 or 500 observations, say, would yield different thresholds, as reported in Simulation~2. The \cite{Wadsworth:2016} method fails on the first two datasets, but can be fixed by selecting just a handful of thresholds. For the Nidd data, the parametric methods select either the highest threshold or nearly all observations.

\begin{table}[ht!]
\centering
\caption{Number of exceedances at chosen threshold for four standard datasets. For parametric methods (bottom), candidate thresholds consisted of 21st, 31st, \ldots, 151st order statistics (Nidd river data), empirical quantiles at level 0.8, 0.81, \ldots, 0.99 (Fort Collins precipitation and Danish insurance data), and empirical quantiles at level 0.5, 0.55, \ldots, 0.90 quantiles (North Sea wave height). Methods with em-dash (---) indicate that no threshold is returned. Stars ($\star$) indicate a procedure run with different or smaller number of thresholds.}\label{tab:real-data}

\begin{tabular}{lrrrr}
\toprule
method & Nidd & Danish & Fort Collins & North Sea\\
\midrule
\cite{Hall.Welsh:1985} & 153 & 19 & 387 & \\
\cite{Hall:1990} & 141 & 81 & 150 & \\
\cite{Beirlant.Vynckier.Teugels:1996b} &  & 770 &  & \\
\cite{Drees.Kaufmann:1998} &  & 64 &  & \\
\cite{Danielsson:2001} & 139 & 1262 & 10 & \\
\cite{Guillou.Hall:2001} & 102 & 84 & 61 & \\
\cite{Dupuis.Victoria-Feser:2003} (non-robust) & 21 & 26 & 10 & \\
\cite{Reiss.Thomas:2007} ($l_1$) & 147 & 1661 & 66 & \\
\cite{Reiss.Thomas:2007} ($l_2$) & 137 & 1547 & 43 & \\
\cite{Goegebeur.Beirlant.deWet:2008} & 145 & 1796 & 699 & \\
\cite{Clauset.Shalizi.Newman:2009} & 89 & 1564 & 586 & \\
\cite{Gomes.Figueiredo.Neves:2012} & 91 & 1285 & 13 & \\
\cite{Gomes:2013} (sample paths) & 140 & 1551 & 419 & \\
\cite{Wager:2014} & 77 & 94 & 2073 & \\
\cite{Caeiro.Gomes:2014} & 86 & 1163 & 11 & \\
\cite{Caeiro.Gomes:2016} (AMSE) & 67 & 546 & 33 & \\
\cite{Danielsson:2019} (eye-balling) & 6 & 21 & 50 & \\
\cite{Danielsson:2019} (MAD) & 12 & 17 & 206 & \\
\cite{Danielsson:2019} (KS) & 16 & 95 & 6 & \\
\cite{Bladt:2020} & 54 & 591 & 318 & \\
\cite{Schneider:2021} (SAMSEE) & 77 & 944 & 59 & \\
 \addlinespace
\cite{Pickands:1975} & 89 & 174 & 362 & 157\\
\cite{Davison.Smith:1990} & 20 & 260 & 810 & 283\\
\cite{Caers.Beirlant.Maes:1999} & 149 & 434 & 737 & 314\\
\cite{Thompson:2009} & 20 & 174 & 162 & 314\\
\cite{Northrop.Coleman:2014} & 40 & 65 & 125 & 95\\
\cite{Langousis:2016} & 149 & 434 & 810 & 157\\
\cite{Castillo.Padilla:2016} & 39 & 174 & 810 & 314\\
\cite{Wadsworth:2016} &  &  & 810 & 157\\
\cite{Northrop.Attalides.Jonathan:2017} & 127 & 434 & 810 & 189\\
\cite{Bader.Yan.Zhang:2018} & 149 & 195 & 39 & 189\\
\cite{SilvaLomba.FragaAlves:2020} & 20 & 22 & 39 & 220\\
\cite{Varty.Tawn:2021} & 149 & 434 & 810 & 314\\
\cite{Kiran.Srinivas:2021} & 133 & 195 & 125 & 157\\
\cite{Gamet.Jalbert:2022} (beta EGP model) &  & 195 & 572 & 189\\
\cite{Murphy.Tawn.Varty:2024} & 30 & 195 &  & 314\\
\bottomrule
\end{tabular}

\end{table}

\section{Omitted methods}

Although some may have escaped our scrutiny or have appeared very recently, we have attempted to describe and compare most threshold selection methods in the literature. Others sketched below were excluded from the simulations for a variety of reasons.

\subsection{Mean residual life plot}

\cite{Minguez:2025} includes additional tests, routinely used in linear regression, for outliers based on internally studentized residuals (for the smallest observation) and goodness-of-fit assessment based on the variance of these residuals. He also suggests fitting smoothing splines to the mean squared error values to identify local minima, but the spline ignores the serial correlation of these values and differences in variance of the point estimates. The multi-stage procedure seems to us overly complex.

\subsection{Parameter stability plots}

\cite{Curceac:2020} propose to fit a a penalized cubic spline to smooth parameter estimates in parameter stability plots until a plateau emerges. The detection of the latter is difficult, and the method does not account for the dependence of successive estimators.

\subsection{Robust methods}
Robust estimation methods can downweight observations that do not conform with the model, thereby reducing their influence on the fit, but comes at the expense of a loss of information \citep[cf.][\S5]{Davison.Smith:1990}. \cite{Dupuis:1999} proposes an optimal $B$-robust estimator of the generalized Pareto parameters, which returns both parameter estimates and weights for observations. In the upper tail, the weights are non-increasing as the quantile increases. Poorly-fitted points are downweighted and thus suspected not to be drawn from the limiting model, but observations from the GPD itself might also be downweighted, so simulation is needed to calibrate the weights. To avoid multiple testing, we used a cusum statistic based on the $m$ non-zero weight of the upper tail fit to the original data, and approximated the null distribution by computing the weights for samples drawn from a GPD using the optimal bias-robust estimator. We get strong indication of missfit for thresholds lower than the 98\% for the Padova data, leading to $u=42.77$mm. The repeated fit is too computationally expensive to consider in a simulation study, and more work on calibration and the choice of summary statistic would be needed for this approach to be properly automated.

\subsection{Coefficient of variation stability}
\cite{Castillo.Padilla:2016} propose using the coefficient of variation for diagnostics and formulate an automated procedure for exceedances of a fixed grid $\calU$ of thresholds. A generalized Pareto random variable $\ddistGP(\sigma, \xi)$ has coefficient of variation $\mathrm{cv}(X) = \mathrm{sd}(X)/\mathrm{E}(X)=(1-2\xi)^{-1/2}$, which does not depend on the scale parameter and yields a consistent moment-based shape estimator if $\xi < 1/2$, which, if $\xi<1/4$ and suitably renormalized, converges in distribution to a Gaussian process with constant mean and known covariance function. Standard results can be then be used to test whether a realisation of this process has a constant mean. \cite{Castillo.Padilla:2016} suggest parametric bootstrap approximations of the distribution of a related statistic, leading to a sequential procedure that starts with the smallest threshold and stops when the bootstrap $p$-value is lower than a level $\alpha$.
They also propose transformations that deal with the effect of the constraint $\xi<1/4$, which limits the applicability of the idea for heavy-tailed data. However, the transformation affects the rate of convergence of the penultimate approximation \citep{Wadsworth/Tawn/Jonathan:2010}: one needs estimates of $\sigma$ and $\xi$, which depend on the threshold, and their methodology requires a preliminary test of the moment condition. Simulation results indicated that this approach is amongst the worst we considered, leading to systematic selection of the lowest possible threshold whenever applicable. We have opted not to discuss it further. It also fails often, and cannot be recommended.

\subsection{Smoothing of Hill's estimator}

\cite{Schneider:2021} use related ideas for the inverse Hill statistic, which approximates the mean integrated squared error of Hill's estimator relative to an exponential density. They use Bayesian nonparametric methods to smooth the resulting estimates and pick $n_u$ for which the smoothed mean is smallest. The empirical Bayes procedure proposed for the smoothing is computationally intensive in medium to large samples, and the degree of smoothing was unsatisfactory in our numerical work.

\end{appendix}

\end{document}